\documentclass[conference]{IEEEtran}
\pdfoutput=1

\IEEEoverridecommandlockouts

\usepackage[normalem]{ulem}
\usepackage[dvipsnames]{xcolor}

\usepackage{pgfmath}
\usepackage{tikz}
\usetikzlibrary{
    calc,
    fadings,
    decorations,
    positioning,
    decorations.markings,
    arrows.meta,
    intersections,
    backgrounds,
    shadows.blur,
    fit,
    decorations.pathreplacing,
    shapes,
    matrix,
    math,
    circuits.logic.US,
    quantikz,
    tikzmark,
    zx-calculus,
    overlay-beamer-styles,
    external
}
\newcommand{\externalize}[2]{%
  \tikzsetnextfilename{#1}%
  {#2}%
  \tikzexternaldisable
}
\newcommand{\externalizezx}[2]{%
  \tikzsetnextfilename{#1}%
  {\tikz{\node[inner sep=0pt, outer sep=0pt]{%
          #2
  };}}%
  \tikzexternaldisable
}

\usepackage{amsmath}
\usepackage{amsthm}
\usepackage{amssymb}
\usepackage{amsfonts}

\usepackage{cite}
\usepackage{algorithmic}
\usepackage{graphicx}
\usepackage{textcomp}

\allowdisplaybreaks[1]
\usepackage{mathtools}
\mathtoolsset{showonlyrefs=true}

\usepackage{iftex}
\usepackage{xfrac}
\usepackage{wrapfig}
\usepackage{xspace}

\usepackage[colorinlistoftodos]{todonotes}
\definecolor{colorZxZ}{gray}{1.0}
\definecolor{colorZxX}{gray}{0.6}
\definecolor{colorZxH}{gray}{1.0}
\tikzset{%
    zxstyle/.style={%
        rounded rectangle,
        anchor=center, 
        line width=0.4pt,
        execute at begin node={\thinmuskip=0mu\medmuskip=0mu\thickmuskip=0mu}, 
        minimum size=3mm,
        font={\fontsize{8}{10}\selectfont\boldmath},
        inner sep=0.0mm,
        inner xsep=1.2mm,
        scale=0.8,
    },
    zxstyletight/.style={%
        zxstyle,
        minimum size=4mm,
        inner ysep=0.2mm,
        inner xsep=0.8mm,
    },
    spider/.style={%
        zxstyle,
        draw=black,
    },
    coordinate/.style={%
        inner sep=0.0pt,
        outer sep=0.0pt,
        anchor=center
    },
    leftsummator/.style={%
        circuit logic US,
        buffer gate,
        draw=black,thin,
        inner sep=0.0ex,
        label={center:\small $\Sigma$},
        anchor=west,
        node contents={}
    },
    rightsummator/.style={%
        circuit logic US,
        buffer gate,
        draw=black,thin,
        rotate=180,
        inner sep=0.0ex,
        label={[rotate=180]center:\small $\Sigma$},
        anchor=west,
        node contents={}
    },
    bubble/.style={%
        ellipse,
        draw=black,thin, 
        inner sep=0.1ex
    },
    xspider/.style={%
        spider,
        fill=gray!70,
        inner sep=#1
    },
    xspider/.default={3pt},
    zspider/.style={%
        spider,
        fill=white,
        inner sep=#1
    },
    zspider/.default={2pt},
    hadamard/.style={%
        rectangle,
        fill=white,
        draw=black,
        anchor=center,
        inner sep=2pt,
        node contents={}
    },
    empty/.style={%
        inner sep=0.0pt,
        outer sep=0.0pt,
        minimum width=0pt,
        minimum height=0pt,
        draw=none,
        anchor=center,
        node contents={}
    },
    gamma/.style={%
        zspider,
        zxstyle,
        inner sep=0.3pt,
        anchor=center,
        node contents={#1}
    },
    gamma/.default={$\gamma$},
    gammatimes2/.style={%
        zspider,
        zxstyle,
        inner sep=0.2pt,
        anchor=center,
        node contents={$2\gamma$}
    },
    gammainv/.style={%
        zspider,
        zxstyle,
        inner sep=0.2pt,
        anchor=center,
        minimum width=0.4em,
        node contents={$\zxMinus$#1}
    },
    gammainv/.default={$\gamma$},
    gammainvtimes2/.style={%
        zspider,
        zxstyle,
        inner sep=0.2pt,
        anchor=center,
        node contents={$\zxMinus2\gamma$}
    },
    beta/.style={%
        xspider,
        zxstyle,
        inner sep=0.3pt,
        anchor=center,
        node contents={#1}
    },
    beta/.default={$\beta$},
    betainv/.style={%
        xspider,
        zxstyle,
        inner sep=0.2pt,
        anchor=center,
        node contents={$\zxMinus$#1}
    },
    betainv/.default={$\beta$},
    betainvtimes2/.style={%
        xspider,
        zxstyle,
        inner sep=0.2pt,
        anchor=center,
        node contents={$\zxMinus2\beta$}
    },
    betatimes2/.style={%
        xspider,
        zxstyle,
        inner sep=0.2pt,
        anchor=center,
        node contents={$2\beta$}
    },
    zident/.style={%
        zspider,
        zxstyle,
        inner sep=0.1pt,
        anchor=center,
        node contents={}
    },
    xident/.style={%
        xspider,
        zxstyle,
        inner sep=0.1pt,
        anchor=center,
        node contents={}
    },
    zpi/.style={%
        zspider,
        zxstyle,
        inner sep=0.4pt,
        anchor=center,
        node contents={$\pi$}
    },
    xpi/.style={%
        xspider,
        zxstyle,
        inner sep=0.4pt,
        anchor=center,
        node contents={$\pi$}
    },
    hedge/.style={%
        dashed,
        color=NavyBlue,
        thick
    }
}

\definecolor{sbblue}{rgb}{0.2823529411764706, 0.47058823529411764, 0.8156862745098039}
\definecolor{sborange}{rgb}{0.9333333333333333, 0.5215686274509804, 0.2901960784313726}
\definecolor{sbgreen}{rgb}{0.41568627450980394, 0.8, 0.39215686274509803}
\definecolor{sbred}{rgb}{0.8392156862745098, 0.37254901960784315, 0.37254901960784315}
\definecolor{sbpurple}{rgb}{0.5843137254901961, 0.4235294117647059, 0.7058823529411765}
\definecolor{sbbrown}{rgb}{0.5490196078431373, 0.3803921568627451, 0.23529411764705882}
\definecolor{sbmagenta}{rgb}{0.8627450980392157, 0.49411764705882355, 0.7529411764705882}
\definecolor{sbgray}{rgb}{0.4745098039215686, 0.4745098039215686, 0.4745098039215686}
\definecolor{sbocca}{rgb}{0.8352941176470589, 0.7333333333333333, 0.403921568627451}
\definecolor{sblightblue}{rgb}{0.5098039215686274, 0.7764705882352941, 0.8862745098039215}

\definecolor{lightgray}{rgb}{0.7,0.7,0.7}
\definecolor{darkgreen}{rgb}{0,0.39,0}
\definecolor{darkblue}{rgb}{0,0,.6}
\definecolor{darkred}{rgb}{.6,0,0}

\newcommand{\ii}{\mathrm{i}}

\newcommand{\ee}{\mathrm{e}}

\newcommand{\SpiderRule}{\ensuremath{({\bm f})}\xspace}

\newcommand{\PiCopyRule}{\ensuremath{({\bm \pi})}\xspace}
\newcommand{\CopyRule}{\ensuremath{({\bm c})}\xspace}
\newcommand{\HadamardRule}{\ensuremath{({\bm{h}})}\xspace}
\newcommand{\IdRule}{\ensuremath{({\bm{id}})}\xspace}
\newcommand{\HHRule}{\ensuremath{({\bm{hh}})}\xspace}
\newcommand{\BialgRule}{\ensuremath{({\bm b})}\xspace}

\newcommand{\HopfRule}{\ensuremath{({\bm{hopf}})}\xspace}

\newcommand{\picopyrule}{{\footnotesize\PiCopyRule}}

\newcommand{\bialgrule}{{\footnotesize\BialgRule}}
\newcommand{\idrule}{{\footnotesize\IdRule}}
\newcommand{\hadamardrule}{{\footnotesize\HadamardRule}}
\newcommand{\hhrule}{\footnotesize\HHRule}
\newcommand{\copyrule}{\footnotesize\CopyRule}

\newcommand{\spiderrule}{\footnotesize\SpiderRule}
\newcommand{\fusionrule}{\footnotesize\SpiderRule}
\newcommand{\hopfrule}{\footnotesize\HopfRule}

\makeatletter
\DeclareRobustCommand{\rvdots}{%
  \vbox{
    \baselineskip4\p@\lineskiplimit\z@
    \kern-\p@
    \hbox{.}\hbox{.}\hbox{.}
  }}
\makeatother

\tikzset{external/system call={lualatex
\tikzexternalcheckshellescape -halt-on-error -interaction=batchmode
-jobname "\image" "\texsource"}}
\tikzexternaldisable

\begin{document}
\title{Measurement-Based Quantum Approximate Optimization}

\author{\IEEEauthorblockN{1\textsuperscript{st} Tobias Stollenwerk}
\IEEEauthorblockA{\textit{Institute for Quantum Computing Analytics (PGI-12)} \\
\textit{J\"ulich Research Center}\\
Wilhelm-Johnen-Str., 
52428 J\"ulich, Germany \\
to.stollenwerk@fz-juelich.de\\
Orcid-ID 0000-0001-5445-8082}
\and
\IEEEauthorblockN{2\textsuperscript{nd} Stuart Hadfield}
\IEEEauthorblockA{\textit{Quantum Artificial Intelligence Lab (QuAIL)} \\
\textit{USRA Research Institute for Advanced Computer Science}\\
\textit{NASA Ames Research Center}, %
CA 94035\\ 
stuart.hadfield@nasa.gov\\
Orcid-ID 0000-0002-4607-3921}
}
\maketitle

\begin{abstract}
Parameterized quantum circuits are attractive candidates for potential quantum advantage in the near term and beyond.  
At the same time, as quantum computing hardware not only continues to improve but also begins to incorporate new features such as mid-circuit measurement and adaptive control, opportunities arise for innovative algorithmic paradigms.
In this work we focus on measurement-based quantum computing protocols for approximate optimization, in particular related to quantum alternating operator ans\"atze (QAOA), 
a popular quantum circuit model approach to combinatorial optimization. 
For the construction and analysis of our measurement-based protocols we demonstrate that diagrammatic approaches, specifically ZX-calculus and its extensions,  
are effective for adapting such algorithms to the measurement-based setting. 
In particular we derive measurement patterns for applying QAOA to the broad and important class of QUBO problems.
We further outline how for constrained optimization,  hard problem constraints may be directly incorporated into our protocol to guarantee the feasibility of the solution found and avoid the need for dealing with penalties.
Finally we discuss the resource requirements and tradeoffs of our approach to that of more traditional quantum circuits. 

\end{abstract}

\section{Introduction}%
There has been much %
interest in quantum approaches to optimization~\cite{abbas2023quantum}, including quantum annealing~\cite{albash2018adiabatic}, and, more recently,
quantum gate model approaches such as quantum alternating operator ans\"atze (QAOA)~\cite{hogg2000quantum,farhi2014quantum,hadfield2019quantum}.
Typically these approaches perform better with increasing circuit depth and decreasing circuit noise, which limits their effectiveness on current or near-term quantum hardware.
Hence there is much recent interest in alternative algorithmic paradigms that use quantum resources in a different way; this includes both different algorithms as well as different computational models.

In this paper we consider
measurement-based quantum computation (MBQC)~\cite{raussendorf2001one,raussendorf2003measurement,jozsa2006introduction,nielsen2006cluster,browne2016one,briegel2009measurement} approaches to tackling hard classical optimization problems. 
While both the quantum circuit- and measurement-based models allow for universal quantum computation~\cite{raussendorf2003measurement}, they are fundamentally different operationally, and present significantly different resource requirements.  
Indeed while gate-model algorithms are limited by the available qubits and number of high-fidelity gates that can be performed, MBQC algorithms are primarily limited by the size of the entangled resource state one can prepare. 
Hence, for some applications the required coherence times or error thresholds may be much less demanding~\cite{raussendorf2003measurement, zwerger2013universal, zwerger2014hybrid, zwerger2016measurement}. 
To this end MBQC is particularly attractive for quantum hardware platforms such as 
photonics~\cite{walther2005experimental,chen2007experimental,schwartz2016deterministic,larsen2019deterministic,zwerger2016measurement} where entangling gates may be challenging to realize but resource states can be prepared probabilistically.

Building on previous work that considered only the QAOA depth $p=1$ (i.e., single-layer) case~\cite{proietti2022native, kaldenbach2023mapping}, 
we show how QAOA may be imported to the measurement-based model for arbitrary depth $p$ and arbitrary algorithm parameters.
To do this we use standard and novel relations between these two models, conveniently derived through a diagrammatic language called ZX-calculus.
While the ZX-calculus has been previously applied to MBQC \cite{duncan2010rewriting,duncan2012graphical,kissinger2019universal,backens2021there}, as well as parameterized quantum circuits~\cite{zhao2021analyzing,stollenwerk2022diagrammatic,koch2022quantum}, to our knowledge our application and approach is novel.
We discuss the resource tradeoffs as compared to circuit-based implementations, 
and further outline how our approach directly extends to the broad and important class of quadratic unconstrained binary optimization (QUBO) problems, 
as well as various constrained optimization problems by importing generalized QAOA mixers that preserve hard problem constraints~\cite{hadfield2019quantum}. 
While general methods to translate gate-based quantum algorithms into the MBQC model exist~\cite{raussendorf2001one,danos2007measurement,browne2016one}, 
they typically come with significant resource overhead and are not guaranteed to deal with different circuit parameters in a uniform way. 
It is hence important to develop flexible application-tailored approaches. 
Though we focus on QAOA our approach serves as a prototype to port other algorithms based on parameterized quantum circuits to the measurement-based paradigm.

The paper is structured as follows. 
We begin by 
briefly overviewing ZX-calculus, QAOA, and the 
MBQC paradigm 
in Section~\ref{sec:preliminaries}. 
In Section~\ref{sec:qaoa_mbqc}, we 
present the conversion of the basic algorithmic elements of QAOA from
the quantum circuit model into a measurement-based protocol, before we  combine these results to formulate measurement-based QAOA for an arbitrary number of layers, 
and discuss the resource requirements thereof. 
While we focus on MaxCut for simplicity, we show how our results generalize to QUBO problems in a straightforward way.  
We next outline how our techniques extend to QAOA with alternative mixing operators that preserve problem constraints by considering the Maximum Independent Set (MIS) problem in Section~\ref{sec:qaoa_mbqc_mis}.  
We give an outlook on further potential extensions of our approach in Section~\ref{sec:outlook} before we conclude in Section~\ref{sec:conclusion}. Along the way several diagrammatic proofs are deferred to an Appendix.

\section{Preliminaries}%
\label{sec:preliminaries}
Here for the reader we overview ZX-calculus, %
measurement-based quantum computing paradigm, and %
QAOA.
\subsection{ZX-Calculus}
Representing quantum circuits in the versatile diagrammatic language called \textit{ZX-calculus} 
turns out to be very useful for transferring quantum algorithms from the gate-model to the measurement based quantum computing paradigm.
In the following, we will introduce the basics of the ZX-calculus. 
We refer the reader to~\cite{vilmart2018near,van2020zx} and the references therein for comprehensive introductions. 
Note that throughout we employ the convention of grayscale notation for ZX-diagrams as used in~\cite{van2020zx} instead of the original red and green diagram colorings.

The ZX-calculus allows translation between 
linear maps, 
such as quantum circuits and expectation values, 
and \textit{string diagrams} that equivalently represent or encode the same object. 
After translation, ZX-diagrams may be 
conveniently rearranged or simplified using a sets of standard, 
mathematically rigorous 
diagram manipulation rules. 
While in the usual circuit diagram representation   
quantum circuits 
correspond to directed acyclic graphs with nodes labeled by a unitary quantum gate, string diagrams are more general 
and hence facilitate a 
potentially more powerful approach.   
In particular, string diagrams correspond to \textit{undirected} graphs, i.e., ZX-diagrams and sub-objects do not inherit the temporal ordering implicit in 
quantum circuit diagrams. 
Hence, while a quantum circuit can always be efficiently translated to an equivalent ZX-diagram, the converse is not true in general~\cite{duncan2020graph}. 
We note that while ZX-diagrams can incorporate the density matrix picture of quantum mechanics, for our purposes it will suffice to consider pure states. 

The basic building blocks of a ZX-diagram are so-called \textit{Z-spiders} and \textit{X-spiders}, denoted as 
\begin{align}
    \begin{ZX}[ampersand replacement=\&]
        \leftManyDots{m} \zxZ{\theta} \rightManyDots{n}
    \end{ZX}
    & := 
    \underbrace{\ket{0\dots 0}}_{n}\underbrace{\bra{0\dots 0}}_m + e^{\ii\theta}\underbrace{\ket{1\dots 1}}_{n}\underbrace{\bra{1\dots 1}}_m %
    \label{eqn:def_z_spider}
    \\
    \begin{ZX}[ampersand replacement=\&]
        \leftManyDots{m} \zxX{\theta} \rightManyDots{n}
    \end{ZX}
    & := 
    \underbrace{\ket{+\dots +}}_{n}\underbrace{\bra{+\dots +}}_m + e^{\ii\theta}\underbrace{\ket{-\dots -}}_{n}\underbrace{\bra{-\dots -}}_m \;, \\ 
    \label{eqn:def_x_spider}
\end{align}
respectively.
Here we employ the common approach of defining various ZX-diagram objects in terms of vectors and matrices in the usual bra-ket notation; 
later it will suffice to consider and manipulate ZX-diagrams alone. 
The numbers of input and output lines $n,m\geq 0$ can be the same or different. 
Roughly speaking, each line corresponds to a qubit, 
though this association may be broken by subsequent diagram manipulations. 
Note that spiders are symmetric tensors and as such are invariant under same-side permutations of wires.

As special cases of~\eqref{eqn:def_z_spider} and~\eqref{eqn:def_x_spider} for a single qubit, the $Z$ and $X$ Pauli operators and their (unnormalized) eigenstates can be represented as 
\begin{align}
    \begin{ZX}
        \zxN{} \rar &[\zxwCol] \zxZ{\pi} \rar & [\zxwCol] \zxN{}
    \end{ZX}
    & = 
    \ket{0}\bra{0} - \ket{1}\bra{1} 
    = 
   Z
    \label{eqn:z_gate_zx}
    \\
    \begin{ZX}
         \zxN{} \rar &[\zxwCol] \zxX{\pi} \rar & [\zxwCol] \zxN{}
    \end{ZX}
    & = 
    \ket{0}\bra{1} + \ket{1}\bra{0}
    = 
   X
    \label{eqn:x_gate_zx}
    \\
    \begin{ZX}
        \zxZ{} \rar & [\zxwCol,\zxwCol] \zxN{}
    \end{ZX}
    & = 
    \ket{0} + \ket{1} 
    = 
    \sqrt{2} \ket{+}
    \label{eqn:plus_state_zx}
    \\
    \begin{ZX}
        \zxX{} \rar & [\zxwCol,\zxwCol] \zxN{}
    \end{ZX}
    & = 
    \ket{+} + \ket{-} 
    = 
    \sqrt{2} \ket{0}
    \label{eqn:zero_state_zx}
    \\
    \begin{ZX}
        \zxZ{\pi} \rar & [\zxwCol] \zxN{}
    \end{ZX}
    & = 
    \ket{0} - \ket{1} 
    = 
    \sqrt{2} \ket{-}
    \label{eqn:minus_state_zx}
    \\
    \begin{ZX}
        \zxX{\pi} \rar & [\zxwCol] \zxN{}
    \end{ZX}
    & = 
    \ket{+} - \ket{-} 
    = 
    \sqrt{2} \ket{1}
    \label{eqn:one_state_zx}
\end{align}
Similarly, single qubit X- and Z-rotations read
\begin{align}
    \begin{ZX}[ampersand replacement=\&]
        \zxN{} \rar \&[1em] \zxZ{\gamma} \rar \&[1em] \zxN{}
    \end{ZX}
    & \propto 
    \vcenter{\hbox{
    \begin{tikzcd}[thin lines, column sep=1em, row sep={4.0ex,between origins}, ampersand replacement=\&]
        \qw \&  \gate{R_Z(\gamma)} \& \qw 
    \end{tikzcd}
    }}
    \label{eqn:zrot_gate_zx}
    \\
    \begin{ZX}[ampersand replacement=\&]
        \zxN{} \rar \&[1em] \zxX{\beta} \rar \&[1em] \zxN{}
    \end{ZX}
    & \propto 
    \vcenter{\hbox{
    \begin{tikzcd}[thin lines, column sep=1em, row sep={4.0ex,between origins}, ampersand replacement=\&]
        \qw \&  \gate{R_X(\beta)} \& \qw 
    \end{tikzcd}
    }}
    \label{eqn:zxot_gate_zx}
\end{align}
Here the notation $\propto$ indicates equal up to constant, which for our purposes is usually an irrelvant global phase or normalization factor.
The Hadamard gate is given a special symbol and denoted 
$H = 
\begin{ZX}
    \zxN{} \rar & [\zxwCol] \zxH{} \rar & [\zxwCol] \zxN{}
\end{ZX}
=
\ee^{-\ii \frac{\pi}{4}}
\begin{ZX}
    \zxN{} \rar &  \zxFracZ{\pi}{2} \rar &  \zxFracX{\pi}{2} \rar &  \zxFracZ{\pi}{2} \rar \rar &  \zxN{}
\end{ZX}
$
.
Other single-qubit gates may be derived similarly by combining these primitives. 

For multi-qubit gates, like the CNOT gate, we have
\begin{equation}
    \text{CNOT} 
    = 
    \begin{quantikz}[baseline=-2pt, thin lines, row sep=3ex, column sep=1em]
        \qw & \ctrl{1} & \qw \\
        \qw & \targ{}  & \qw
    \end{quantikz}
    =
    \sqrt{2}
    \begin{ZX}
        \zxN{} \rar &[\zxwCol] \zxZ{} \dar \rar & [\zxwCol] \zxN{} 
        \\[2ex]
        \zxN{} \rar &[\zxwCol] \zxX{}      \rar & [\zxwCol] \zxN{}
    \end{ZX}
    \label{eqn:cnot_gate_zx} \; ,
\end{equation}
and the CZ gate reads
\begin{equation}
    \text{CZ} 
    = 
    \begin{quantikz}[baseline=-2pt, thin lines, row sep=3ex, column sep=1em]
        \qw & \ctrl{1}  & \qw \\
        \qw & \ctrl{-1} & \qw
    \end{quantikz}
    =
    \sqrt{2}
    \begin{ZX}
        \zxN{} \rar &[\zxwCol] \zxZ{} \ar[d, H] \rar & [\zxwCol] \zxN{} 
        \\[2ex]
        \zxN{} \rar &[\zxwCol] \zxZ{}           \rar & [\zxwCol] \zxN{}
    \end{ZX} \; .
    \label{eqn:cz_gate_zx}
\end{equation}
Hence it is immediate that together these %
primitives 
can represent a universal set of quantum gates.

\begin{figure}[htpb]
    \centering
    \externalize{zx_rules}{%
        \begin{tikzpicture}
            \input{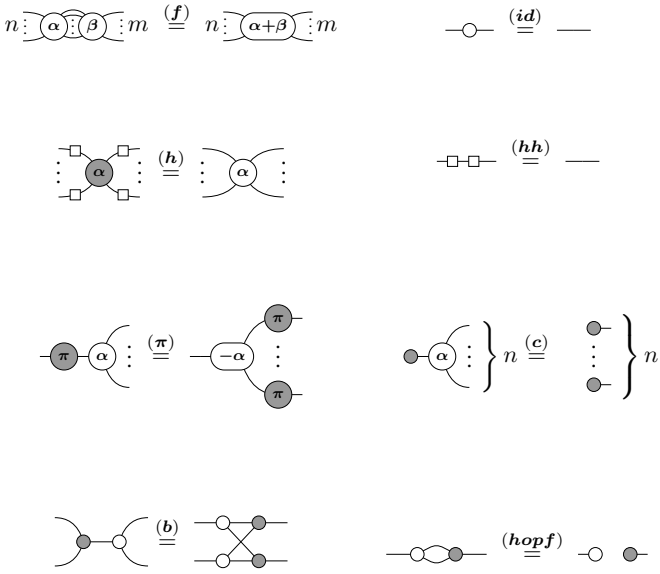}
        \end{tikzpicture}
    }
    \caption{The ZX-diagram rewrite rules (cf.\ for example~\cite{van2020zx} or~\cite{zhao2021analyzing}).}%
    \label{fig:zx_rules}
\end{figure}

We next explain how to manipulate ZX-diagrams.
After translating a quantum circuit to a ZX-diagram, we are left with an undirected graph consisting of internal nodes (Z- and X-spiders), as well as input and output nodes (cf.~\cite{duncan2020graph}). 
As long as the input and output nodes are kept fixed, the position of the internal nodes can be arbitrarily manipulated as long as the topology stays fixed. 
In addition, there are a number ZX-diagram rewrite rules which are displayed in Figure~\ref{fig:zx_rules}, which can be used to transform the topology and type of the internal nodes in an equivalent way.
We use the label attached to each equation to reference these rules when we apply them in the examples we consider below. 
While diagram rewriting can be employed towards the simplification of quantum circuits, as we will demonstrate shortly, it can also be used to transform between quantum circuit and measurement-based protocols.

\subsection{Measurement-based Quantum Computing}
In the gate-model a unitary transformation, usually represented as sequence of single- and two-qubit-gates (the quantum circuit) is applied to an initial state which usually assumed to be a disentangled product state.
Although this is the standard way to describe quantum algorithms, there is an equivalent model for quantum computation, the measurement-based quantum computation (MBQC) or one-way quantum computing model \cite{raussendorf2001one,raussendorf2003measurement,jozsa2006introduction,nielsen2006cluster,browne2016one,briegel2009measurement}. 
In contrast to the gate-model, in the MBQC paradigm, the initial state is a highly entangled \emph{resource state} or \emph{graph state},
that is usually given by
\begin{equation}
    \ket{G} = \prod_{(u, v) \in E} \mathrm{CZ}_{u, v} \ket{+}^{\otimes |V|}
    \label{eq:graph_state}
\end{equation}
where the graph $G=(V, E)$ with vertices $V$ and edges $E$, is usually a planar one due to hardware constraints~\cite{briegel2009measurement}.
The quantum algorithm is then performed solely by single-qubit operations (gates and measurements) on the qubits of the graph state.
Note, that the graph state is therefore usually independent of the algorithm.

The graph state has a natural representation in ZX-calculus.
To illustrate, consider the simple square graph with edge set $E=\{(1, 2), (2, 3), (3, 4), (4, 1)\}$ and 4 vertices.
The graph state is given in the gate-model as
\begin{align}
    \ket{G} 
    &= \prod_{(u, v) \in E} \mathrm{CZ}_{u, v} \ket{+}^{\otimes |V|} 
    =
    \vcenter
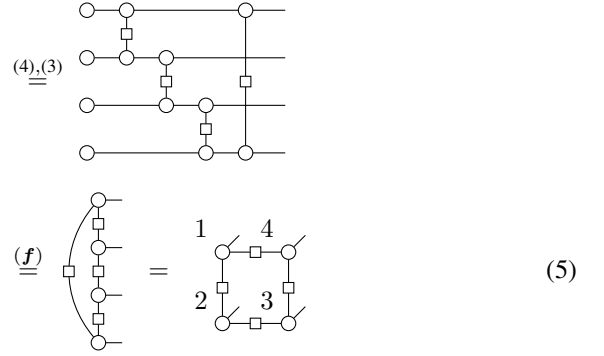
{\hbox{
        \begin{tikzcd}[thin lines, row sep=2.2ex, column sep=0.3em]
    \lstick{\ket{+}} & \qw & \ctrl{1}  & \qw & \qw       & \qw & \qw       & \qw & \ctrl{3}  & \qw & \qw \\
    \lstick{\ket{+}} & \qw & \ctrl{-1} & \qw & \ctrl{1}  & \qw & \qw       & \qw & \qw       & \qw & \qw \\
    \lstick{\ket{+}} & \qw & \qw       & \qw & \ctrl{-1} & \qw & \ctrl{1}  & \qw & \qw       & \qw & \qw \\
    \lstick{\ket{+}} & \qw & \qw       & \qw & \qw       & \qw & \ctrl{-1} & \qw & \ctrl{-3} & \qw & \qw
\end{tikzcd}

    }}
    \\[2ex]
    &
    \stackrel{\eqref{eqn:cz_gate_zx}, \eqref{eqn:plus_state_zx}}{=}
    \vcenter{\hbox{
        \externalize{mbqcexamplediagram00}{%
            \begin{tikzpicture}[node distance=4ex and 1.5em]
                \node[on grid, name=1-l, zident];
\node[on grid, name=2-l, below=of 1-l, zident];
\node[on grid, name=3-l, below=of 2-l, zident];
\node[on grid, name=4-l, below=of 3-l, zident];

\node[on grid, name=1-e12, right=of 1-l, zident];
\node[on grid, name=2-e12, right=of 2-l, zident];
\node[on grid, name=3-e12, right=of 3-l, empty];
\node[on grid, name=4-e12, right=of 4-l, empty];
\draw (1-e12) -- (2-e12) node[midway, hadamard];

\node[on grid, name=1-e23, right=of 1-e12, empty];
\node[on grid, name=2-e23, right=of 2-e12, zident];
\node[on grid, name=3-e23, right=of 3-e12, zident];
\node[on grid, name=4-e23, right=of 4-e12, empty];
\draw (2-e23) -- (3-e23) node[midway, hadamard];

\node[on grid, name=1-e34, right=of 1-e23, empty];
\node[on grid, name=2-e34, right=of 2-e23, empty];
\node[on grid, name=3-e34, right=of 3-e23, zident];
\node[on grid, name=4-e34, right=of 4-e23, zident];
\draw (3-e34) -- (4-e34) node[midway, hadamard];

\node[on grid, name=1-e41, right=of 1-e34, zident];
\node[on grid, name=2-e41, right=of 2-e34, empty];
\node[on grid, name=3-e41, right=of 3-e34, empty];
\node[on grid, name=4-e41, right=of 4-e34, zident];
\draw (4-e41) -- (1-e41) node[midway, hadamard];

\node[on grid, name=1-r, right=of 1-e41, empty];
\node[on grid, name=2-r, right=of 2-e41, empty];
\node[on grid, name=3-r, right=of 3-e41, empty];
\node[on grid, name=4-r, right=of 4-e41, empty];

\begin{scope}[on background layer]
    \draw (1-l) -- (1-r);
    \draw (2-l) -- (2-r);
    \draw (3-l) -- (3-r);
    \draw (4-l) -- (4-r);
\end{scope}

            \end{tikzpicture}
        }
    }}
    \\[2ex]
    &
    \stackrel{\spiderrule}{=}
    \vcenter{\hbox{
        \externalize{mbqcexamplediagram01}{%
            \begin{tikzpicture}[node distance=4ex and 1.5em]
                \node[on grid, name=1, zident];
\node[on grid, name=2, below=of 1, zident];
\node[on grid, name=3, below=of 2, zident];
\node[on grid, name=4, below=of 3, zident];

\draw (1) -- (2) node[midway, hadamard];
\draw (2) -- (3) node[midway, hadamard];
\draw (3) -- (4) node[midway, hadamard];
\draw (4) to[bend left=40]  node[midway, hadamard]{}(1);

\draw (1) -- ++(0:2ex);
\draw (2) -- ++(0:2ex);
\draw (3) -- ++(0:2ex);
\draw (4) -- ++(0:2ex);

            \end{tikzpicture}
        }
    }}
    =
    \vcenter{\hbox{
        \externalize{mbqcexamplediagram02}{%
            \begin{tikzpicture}[node distance=6ex and 2.5em]
                \node[on grid, name=1, zident, label=135:$1$];
\node[on grid, name=2, below=of 1, zident, label=135:$2$];
\node[on grid, name=3, right=of 2, zident, label=135:$3$];
\node[on grid, name=4, above=of 3, zident, label=135:$4$];

\draw (1) -- (2) node[midway, hadamard];
\draw (2) -- (3) node[midway, hadamard];
\draw (3) -- (4) node[midway, hadamard];
\draw (4) -- (1) node[midway, hadamard];

\draw (1) -- ++(45:2ex);
\draw (2) -- ++(45:2ex);
\draw (3) -- ++(45:2ex);
\draw (4) -- ++(45:2ex);

            \end{tikzpicture}
        }
    }}%
    \label{eqn:graph_state_example}
\end{align}
In the last step, we made use of the fact that only topology matters and we can freely change the position of the nodes as long as we keep track of the four outputs of the four-qubit state.
This procedure can be extended to arbitrary graph states, where the resulting ZX-diagram has the same structure as the original graph.

The quantum algorithm is then
specified
by a so-called \emph{measurement pattern} 
which consist of a 
sequence of 
adaptively chosen 
single qubit measurements and 
corrective single qubit gates 
applied to the algorithm-independent
resource state~\cite{raussendorf2003measurement,briegel2009measurement,duncan2012graphical}.
The measurement pattern
must be deterministic, which means that each measurement
can only depend on measurement outcomes from 
earlier in the sequence.
This can be formalized as a flow condition for the measurement patterns~\cite{danos2006determinism,browne2007generalized} that have a natural graphical representation in ZX-calculus~\cite{duncan2012graphical}.
We denote measurement patterns as follows: 
$\mathcal{M}^i_P\to n$
for a projective measurement of qubit $i$ in Pauli basis $P\in\{X, Y, Z\}$ and storing the result in the binary variable $n\in\{0, 1\}$, and
$\Lambda^i_n(U)$
for applying a single qubit gate $U$ onto qubit $i$ if and only if the binary variable $n=1$.
As an example, consider the sequence of measurement patterns
\begin{equation*}
    \left\{%
        \mathcal{M}^4_Z\to n,%
        \mathcal{M}^2_X\to m, %
        \Lambda^3_m(X) %
    \right\}
\end{equation*}
onto the graph state \eqref{eqn:graph_state_example} 
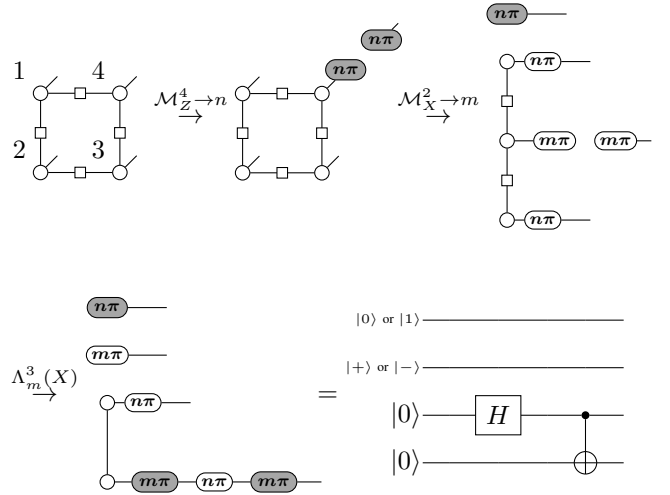
\begin{align*}
    &%
    \vcenter{\hbox{%
            \begin{tikzpicture}[node distance=3.0em and 3.0em]%
                \node[on grid, name=1, zident, label=135:$1$];
\node[on grid, name=2, below=of 1, zident, label=135:$2$];
\node[on grid, name=3, right=of 2, zident, label=135:$3$];
\node[on grid, name=4, above=of 3, zident, label=135:$4$];

\draw (1) -- (2) node[midway, hadamard];
\draw (2) -- (3) node[midway, hadamard];
\draw (3) -- (4) node[midway, hadamard];
\draw (4) -- (1) node[midway, hadamard];

\draw (1) -- ++(45:2ex);
\draw (2) -- ++(45:2ex);
\draw (3) -- ++(45:2ex);
\draw (4) -- ++(45:2ex);

            \end{tikzpicture}
    }} \notag
    \stackrel{\mathcal{M}^4_Z\to n}{\to}%
    \vcenter{\hbox{%
        \externalize{mbqcexamplediagram03}{%
            \begin{tikzpicture}[node distance=3.0em and 3.0em]%
                \node[on grid, name=1, zident];
\node[on grid, name=2, below=of 1, zident];
\node[on grid, name=3, right=of 2, zident];
\node[on grid, name=4, above=of 3, zident];

\draw (1) -- (2) node[midway, hadamard];
\draw (2) -- (3) node[midway, hadamard];
\draw (3) -- (4) node[midway, hadamard];
\draw (4) -- (1) node[midway, hadamard];

\node[on grid, name=m4, xshift=2ex, above=2ex of 4, xspider] {$n \pi$};
\draw (1) -- ++(45:2ex);
\draw (2) -- ++(45:2ex);
\draw (3) -- ++(45:2ex);
\draw (4) -- (m4);

\node[on grid, name=m4c, xshift=3ex, above=2.5ex of m4, xspider] {$n \pi$};
\draw (m4c) -- ++(45:2ex);

            \end{tikzpicture}
        }
        \vspace*{2.5ex}
        \hspace*{-1.5em}
    }} \notag
    \stackrel{\mathcal{M}^2_X\to m}{\to}%
    \vcenter{\hbox{%
        \externalize{mbqcexamplediagram08}{%
            \begin{tikzpicture}[node distance=3.0em and 3.0em]%
                \node[on grid, name=1, zident];
\node[on grid, name=2, below=of 1, zident];
\node[on grid, name=3, below=of 2, zident];

\draw (1) -- (2) node[midway, hadamard];
\draw (2) -- (3) node[midway, hadamard];
\node[on grid, name=m2, right=4ex of 2, zspider] {$m \pi$};
\draw (2) -- (m2);
\node[name=m2c, right=1.5em of m2, zspider] {$m \pi$};
\draw (m2c) -- ++(0:3ex);

\node[on grid, name=1r, right=3ex of 1, zspider]{$n\pi$};
\node[on grid, name=3r, right=3ex of 3, zspider]{$n\pi$};
\draw (1) -- (1r);
\draw (1r) -- ++(0:4ex);
\draw (3) -- (3r);
\draw (3r) -- ++(0:4ex);

\node[on grid, name=4, above=4ex of 1, xspider] {$n \pi$};
\draw (4) -- ++(0:5ex);

            \end{tikzpicture}
        }
    }} \notag
    \\[5ex]
    &%
    \stackrel{\Lambda^3_m(X)}{\to}%
    \vcenter{\hbox{%
        \externalize{mbqcexamplediagram12}{%
            \begin{tikzpicture}[node distance=3.0em and 3.0em]%
                \node[on grid, name=1, zident];
\node[on grid, name=3, below=of 1, zident];
\draw (1) -- (3);

\node[on grid, name=1r, right=3ex of 1, zspider]{$n\pi$};
\node[on grid, name=3r, right=4ex of 3, xspider] {$m\pi$};
\node[on grid, name=3rr, right=5ex of 3r, zspider]{$n\pi$};
\node[on grid, name=3rrr, right=5ex of 3rr, xspider]{$m\pi$};
\draw (1) -- (1r);
\draw (1r) -- ++(0:4ex);
\draw (3) -- (3r);
\draw (3r) -- (3rr);
\draw (3rr) -- (3rrr);
\draw (3rrr) -- ++(0:4ex);

\node[on grid, name=2, above=4ex of 1, zspider] {$m\pi$};
\draw (2) -- ++(0:5ex);
\node[on grid, name=4, above=4ex of 2, xspider] {$n\pi$};
\draw (4) -- ++(0:5ex);

            \end{tikzpicture}
        }
        \hspace*{-1.5em}
    }} \notag
    =%
    \vcenter{\hbox{%
        \externalizezx{mbqcexamplecircuitfinal}{%
            \begin{tikzcd}[thin lines, row sep={4ex,between origins}, column sep=1.0em, ampersand replacement=\&]
    \lstick{\tiny \ket{0} \text{or} \ket{1}} \& \qw \& \qw       \& \qw \& \qw       \& \qw \&  \\
    \lstick{\tiny \ket{+} \text{or} \ket{-}} \& \qw \& \qw       \& \qw \& \qw       \& \qw \&  \\
    \lstick{\ket{0}}                         \& \qw \& \gate{H}  \& \qw \& \ctrl{1}  \& \qw \&  \\
    \lstick{\ket{0}}                         \& \qw \& \qw       \& \qw \& \targ{}   \& \qw \& 
\end{tikzcd}

        }
    }} \notag
\end{align*}
which leads to the creation of a Bell state in qubits $1$ and $3$. The complete derivation using ZX-calculus is given in  Appendix~\ref{sec:proof_mbqc_example}.
Below in Section~\ref{sec:qaoa_mbqc} we will see further examples of such measurement patterns described by ZX-diagrams.

\subsection{QAOA}
Parameterized quantum circuits (PQC) have gained much attention in recent years, in particular as heuristic approaches suitable for NISQ~\cite{preskill2018quantum} era devices that are classically optimized  %
as part of a hybrid 
protocol, though we emphasize they are by no means restricted to this setting; see \cite{cerezo2021variational,bharti2022noisy,abbas2023quantum} for reviews of recent developments.  
Two particular approaches of interest are the QAOA~\cite{hadfield2019quantum}, which generalizes the quantum approximate optimization algorithm~\cite{farhi2014quantum}) and VQE (variational quantum eigensolver~\cite{peruzzo2014variational,mcclean2016theory}) paradigms, 
as well as a number of more recent variants of these approaches. 
Here we briefly review the original QAOA paradigm and its application to combinatorial optimization, though our results to follow may be applied more generally to a variety of problems and algorithms. 

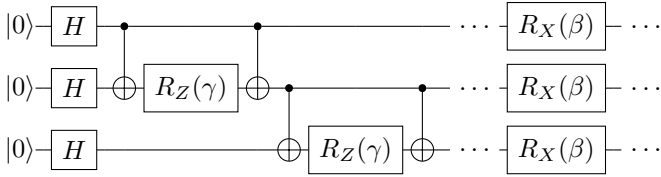
\begin{figure}[htpb]
    \centering
    \begin{quantikz}[thin lines, row sep=1.2ex, column sep=0.3em]
          \ket{0} & \qw & \gate{H} & \qw & \ctrl{1} & \qw                
        & \ctrl{1}  & \qw      &    \qw             &  \qw     & \qw 
        & \ \ldots\ \qw & \gate{R_X(\beta)} & \qw & \ \ldots\  \qw \\
         \ket{0} & \qw & \gate{H} & \qw & \targ{}  
        & \gate{R_Z(\gamma)} & \targ{}   & \ctrl{1}  &    \qw             & \ctrl{1} & \qw 
        & \ \ldots\ \qw & \gate{R_X(\beta)} & \qw & \ \ldots\  \qw \\
        \ket{0} & \qw & \gate{H} & \qw & \qw      & \qw               
        &   \qw   & \targ{}  & \gate{R_Z(\gamma)} & \targ{}    & \qw 
        & \ \ldots\ \qw & \gate{R_X(\beta)} & \qw & \ \ldots\  \qw 
    \end{quantikz}
    \caption{Example of a parameterized quantum circuit: QAOA on 3 qubits. Here the phase and mixing operators as well as initial state preparation have been compiled to basic quantum gates.}%
    \label{fig:example_pqc}
\end{figure}

In QAOA we are given a real cost function~$c(x)$ we seek to optimize over bit strings $x\in\{0,1\}^n$, and corresponding classical Hamiltonian $C$ (i.e., diagonal in the computational basis $C=a_0I + \sum_j a_j Z_j + \sum_{i<j}a_{ij}Z_iZ_j+\dots$ with $C\ket{x}=c(x)\ket{x}$~\cite{hadfield2021representation}). %
We also require specification of a suitable initial state~$\ket{s}$ and parameterized mixing operator~$U_M(\beta)$. 
A QAOA$_p$ circuit then 
consists of $2p$ alternating layers  
\begin{equation}
    \ket{\boldsymbol{\gamma \beta}}= U_M(\beta_p)U_P(\gamma_p)\dots U_M(\beta_1)U_P(\gamma_1)\ket{s} \, ,
\end{equation}
of the mixer and phase operator $U_P(\gamma)=\exp(-\ii \gamma C)$. Here we consider the standard initial state $\ket{s}=\ket{+}^{\otimes n}$ and (transverse-field) mixing operator~$U_M(\beta)=\exp(-\ii \beta B)$, where $B=\sum_{i=1}^n X_i$, %
as originally propsed in ~\cite{farhi2014quantum}.  
Figure~\ref{fig:example_pqc} shows a simple example of a QAOA circuit.
Each QAOA$_p$ state is specified by $2p$ parameters $\gamma_i,\beta_i$ in some domain (e.g.,~$[-\pi,\pi]$), which can be optimized variationally or by other means. Importantly, it can be shown under mild conditions that QAOA performance generally improves with increasing number of layers~$p$~\cite{farhi2014quantum}. 

After preparing on the quantum computer, 
the QAOA state is measured in the computational basis 
which returns some $y\in\{0,1\}^n$ achieving cost $c(y)$. 
Repeated state preparation and measurement gives further samples which may be used to estimate the cost expectation $\langle C\rangle_p$ or other important quantities. 
In particular these quantities could be used to update or variationally search for better circuit parameters if desired. Clearly, algorithm performance will depend on the finding high-quality circuit parameters to begin with, and we emphasize that in different cases parameters may be found through analytic~\cite{wang2018quantum}, numeric~\cite{farhi2014quantum}, or average-case~\cite{streif2019training} techniques, among others, %
though parameter setting remains challenging generally. 
Finally, after a set number of overall runs, or when other suitable termination criteria has been reached, the best overall solution found is returned.

We next turn to translating QAOA circuits into measurement-based protocols via the ZX-calculus. We remark that while some recent work~\cite{proietti2022native,ferguson2021measurement,kaldenbach2023mapping,qin2023applicability,schroeder2023deterministic,chan2023hybrid,mantilla2023measurement,majumder2023variational} 
has begun to consider adaptation of parameterized quantum circuits and variational quantum algorithms to measurement-based settings, as explained, our applications to arbitrary depth QAOA and techniques based on ZX-calculus are novel and hopefully helpful toward further research progress here. We further emphasize that while measurement-based implementations allow a different allocation of quantum resources, high-level algorithmic challenges remain such as parameter setting.

\section{QAOA in the MBQC paradigm}%
\label{sec:qaoa_mbqc}
In this section we show how to transform QAOA from its original formulation in the gate-model into the measurement-based quantum computing paradigm via ZX-calculus.
Here we consider the general class of QUBO problems which includes a wide variety of challenging optimization problems~\cite{lucas2014ising,hadfield2021representation}. (A prototypical example to keep in mind is the MaxCut problem, where given a graph $G=(V,E)$ on $n$ vertices we seek a partition of the vertices that maximizes the number of crossing edges, which is easily shown to be represented by the cost Hamiltonian $C=\tfrac{|E|}2I -\tfrac12\sum_{(ij)\in E} Z_iZ_j$.) Nevertheless it is straightforward to extend our constructions here to QAOA for higher-order problems beyond quadratic. 
Note that
constant factors in the QUBO cost Hamiltonian can be absorbed into the $\gamma,\beta$ parameters as needed and so can be ignored for our constructions to follow.

We begin by considering the QAOA phase-separation operator. 
As the terms of the cost Hamiltonain mutually commute, the phase-separation operator %
can be expressed as a product of 
single qubit rotation gates
\begin{equation}
    \ee^{\ii \gamma Z_u}
    \propto    
    \begin{ZX}[ampersand replacement=\&]
        \zxN{} \rar \&[1em] \zxZ{\gamma} \rar \&[1em] \zxN{}
    \end{ZX} %
    \label{eqn:single_z_rot}
    \; ,
\end{equation}
and two-qubit rotations which correspond to the interaction graph of the problem Hamiltonian, and can be decomposed using phase-gadgets as 
\begin{align}
    \prod_{(u, v) \in E} \ee^{\ii \gamma Z_u Z_v}
    &=
    \prod_{(u, v) \in E} %
    \vcenter{\hbox{%
        \vspace*{-1ex}
            \begin{tikzcd}[thin lines, column sep=1em, row sep={4.0ex,between origins}, ampersand replacement=\&]
    \qw \& \ctrl{1} \& \qw                \& \ctrl{1} \& \qw \\
    \qw \& \targ{}  \& \gate{R_Z(\gamma)} \& \targ{}  \& \qw 
\end{tikzcd}
    }}
    \\[2ex]
    &=
    \prod_{(u, v) \in E}
    \vcenter{\hbox{%
        \vspace*{-3ex}
            \begin{ZX}[row sep={2.5ex}, column sep=1em, ampersand replacement=\&]
    \zxN{} \rar \& \zxZ{} \rar      \&[1em] \zxN{} \rar       \& \zxN{}\\
                \& \zxX{} \uar      \&[1em] \zxZ{\gamma} \lar \&       \\
    \zxN{} \rar \& \zxZ{} \uar \rar \&[1em] \zxN{} \rar       \& \zxN{}\\
\end{ZX}

    }}.%
    \label{eqn:phase_gadget_zx} 
\end{align}
The proof can be found in Appendix~\ref{sec:proof_phase_gadget_zx}.
We can utilize the ZX-rewrite rules~(Figure~\ref{fig:zx_rules}) to transform the phase-gadget towards an MBQC-like formulation
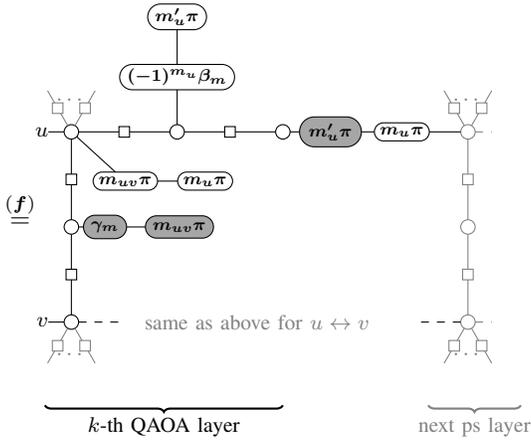
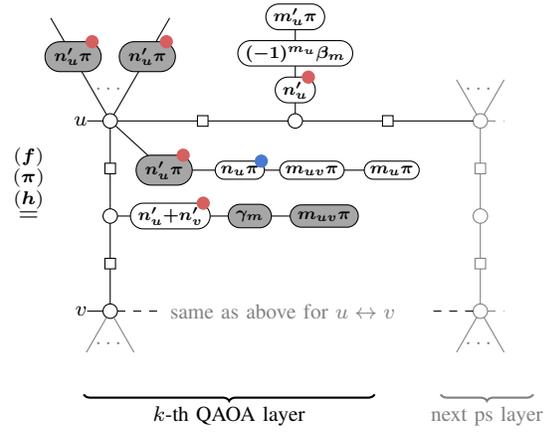
\begin{align}
    \vcenter{\hbox{%
        \externalizezx{phasegadgetdiagram07}{%
            \begin{ZX}[row sep={2.5ex}, column sep=1em, ampersand replacement=\&]
    \zxN{} \rar \& \zxZ{} \rar      \&[1em] \zxN{} \rar       \& \zxN{}\\
                \& \zxX{} \uar      \&[1em] \zxZ{\gamma} \lar \&       \\
    \zxN{} \rar \& \zxZ{} \uar \rar \&[1em] \zxN{} \rar       \& \zxN{}\\
\end{ZX}

        }
    }}
    & \stackrel{\hadamardrule}{=}%
    \vcenter{\hbox{%
        \externalizezx{phasegadgetdiagram10}{%
            \begin{ZX}[zx column sep=1em, ampersand replacement=\&]
    \zxN{} \rar \&[2em] \zxN{} \rar \& \zxZ{} \rar      \& \zxN{} \rar       \& \zxN{} \\
                \&[2em]             \& \zxH{} \uar      \&                   \&        \\
                \&[2em] \zxZ{} \rar \& \zxZ{} \uar \rar \& \zxX{\gamma} \rar \& \zxX{} \\
                \&[2em]             \& \zxH{} \uar      \&                   \&        \\
    \zxN{} \rar \&[2em] \zxN{} \rar \& \zxZ{} \uar \rar \& \zxN{} \rar       \& \zxN{} \\
\end{ZX}

        }
    }}
    \\[2ex]
    & \stackrel{\spiderrule}{=}%
    \vcenter{\hbox{%
        \externalizezx{phasegadgetdiagram12}{%
            \begin{ZX}[zx column sep=1em, ampersand replacement=\&]
    \zxN{} \rar \&[2em] \zxN{} \rar \& \zxZ{} \rar      \& \zxN{} \rar       \& \zxN{}          \& \zxN{}      \\
                \&[2em]             \& \zxH{} \uar      \&                   \&                 \&             \\
                \&[2em] \zxZ{} \rar \& \zxZ{} \uar \rar \& \zxX{\gamma} \rar \& \zxX{m\pi} \rar \& \zxX{m\pi} \\
                \&[2em]             \& \zxH{} \uar      \&                   \&                 \&             \\
    \zxN{} \rar \&[2em] \zxN{} \rar \& \zxZ{} \uar \rar \& \zxN{} \rar       \& \zxN{}          \& \zxN{}      \\
\end{ZX}

        }
    }}
    \\[2ex]
    & \stackrel{\substack{\spiderrule \\ \hadamardrule \\ \picopyrule}}{=}%
    \vcenter{\hbox{%
        \externalizezx{phasegadgetdiagram16}{%
            \begin{ZX}[zx column sep=1em, ampersand replacement=\&]
    \zxN{} \rar \&[2em] \zxN{} \rar \& \zxZ{}       \rar      \& \zxN{} \rar       \& \zxZ{m\pi} \rar       \& \zxN{}      \\
                \&[2em]             \& \zxH{}       \uar      \&                   \&                       \&             \\
                \&[2em] \zxZ{} \rar \& \zxZ{}       \uar \rar \& \zxN{} \rar       \& \zxX{\gamma} \rar     \& \zxX{m\pi} \\
                \&[2em]             \& \zxH{}       \uar      \&                   \&                       \&             \\
    \zxN{} \rar \&[2em] \zxN{} \rar \& \zxZ{}       \uar \rar \& \zxN{} \rar       \& \zxZ{m\pi} \rar       \& \zxN{}      \\
\end{ZX}
        }
    }} %
    \label{eqn:phase-gadget-to-mbqc}
\end{align}
where we have introduced an identity 
$%
\zx{\zxN{} \rar &[1ex] \zxX{2\pi m} \rar &[1ex] \zxN{}}%
=\zx{\zxN{} \rar &[2em] \zxN{}}%
$,
with $m\in\{0, 1\}$, which we identify with the measurement outcome of an ancilla qubit for each edge $(u, v)$ in the basis $\{\ket{0}, \ket{1}\}$, previously initialized in $\ket{+}$.
The complete derivation can be found in Appendix~\ref{sec:proof_phase_gadet_to_mbqc}.
With a similar strategy, we can transform the mixing operator for each node $v \in V$ as 
\begin{align}
    \ee^{\ii \beta X_v}
    &\propto
    \vcenter{\hbox{%
        \externalizezx{mixerdiagram00}{%
            \tikzstyle{basicshadow}=[blur shadow={shadow blur steps=8, shadow xshift=0.7pt, shadow yshift=-0.7pt, shadow scale=1.02}]
\tikzstyle{basic}=[draw,fill=white,basicshadow]
\tikzstyle{operator}=[basic,minimum size=1.5em]

\begin{tikzcd}[thin lines, column sep=1em, row sep={4.0ex,between origins}, ampersand replacement=\&]
    \qw \&  \gate{R_X(\beta)} \& \qw 
\end{tikzcd}
        }
    }} \notag
    \\[2ex] 
    &\propto
    \vcenter{\hbox{%
        \externalizezx{mixerdiagram01}{%
            \begin{ZX}[ampersand replacement=\&]
    \zxN{} \rar \&[1em] \zxX{\beta} \rar \&[1em] \zxN{}
\end{ZX}
        }
    }}\notag
    \\[2ex]
    &=%
    \vcenter{\hbox{%
        \externalizezx{mixerdiagram09}{%
            \begin{ZX}[zx row sep=4ex, zx column sep=1em, ampersand replacement=\&]
    \zxN{} \rar \&[2em] \zxN{} \rar \& \zxZ{}  \rar           \& \zxZ{m_v \pi}                \&                      \&        \\
                \&[2em] \zxZ{} \rar \& \zxZ{}  \ar[u, H] \rar \& \zxZ{(-1)^{m_v}\beta} \rar   \& \zxZ{m'_v \pi}       \&        \\
                \&[2em] \zxZ{} \rar \& \zxZ{}  \ar[u, H] \rar \& \zxX{m'_v \pi} \rar          \&  \zxZ{m_v \pi} \rar \& \zxN{} \\
\end{ZX}
        }
    }} %
    \label{eqn:mixer-to-mbqc} %
    \,.
\end{align}
Here we have introduced two ancilla qubits for each node $v$ with the corresponding measurement outcomes $m_v$ and $m'_v$ in the basis $\{\ket{+}, \ket{-}\}$.
Note, that the input qubit is measured and the information is transferred to the second ancilla qubit.
For a complete derivation see Appendix~\ref{sec:proof_mixer_in_mbqc}.
Similarly we can derive the MBQC version of a Z rotation Eq.~\eqref{eqn:single_z_rot} 
\begin{equation}
    \ee^{\ii \gamma Z_v}
    \propto%
    \vcenter{\hbox{%
        \externalizezx{zrotationdiagram01}{%
            \begin{ZX}[ampersand replacement=\&]
    \zxN{} \rar \&[1em] \zxZ{\gamma} \rar \&[1em] \zxN{}
\end{ZX}
        }
    }} %
    =%
    \vcenter{\hbox{%
        \externalizezx{zrotationdiagram03}{%
            \begin{ZX}[ampersand replacement=\&]
    \zxN{} \rar \& \zxN{} \rar \& \zxZ{} \ar[d, H] \rar \& \zxZ{\gamma}    \rar \& \zxZ{m\pi}      \& \zxN{} \\[3ex]
    \zxN{}      \& \zxZ{} \rar \& \zxZ{} \rar           \& \zxZ{m\pi} \ar[r]    \& \zxN{}     \rar \& \zxN{}  
\end{ZX}
        }
    }} %
    \label{eqn:zrotation_in_mbqc} \; ,
\end{equation}
where we introduced a single ancilla and a measurement with outcome $m$ in the basis $\{\ket{+}, \ket{-}\}$ on the input qubit.
The complete derivation can be found in Appendix~\ref{sec:proof_zrotation_in_mbqc}.
Obviously all the above measurement outcomes are uses for corrections in a causal fashion so that deterministic measurement patterns can be constructed.

Now we can put things together and move towards a MBQC description of QAOA.
For the moment we will
neglect the single qubit Z-rotations in the problem Hamiltonian \eqref{eqn:single_z_rot} for simplicity.   
We will discuss how to extend to arbitrary QUBO problems below.
Consider the initial state $\ket{+}^{\otimes n}$ and the first 
QAOA layer 
\begin{align*}
    &
    \displaystyle
    \prod_{v\in V}\ee^{\ii \beta X_v}%
    \prod_{(u, v) \in E} \ee^{\ii \gamma Z_u Z_v}%
    \ket{+}^{\otimes |V|} =
    \\
    &
    \vcenter{\hbox{%
        \externalizezx{qaoambqcdiagram00}{%

        }
    }}
\end{align*}
were we have depicted only the contribution from a single edge $(u, v)$ for simplicity.
In order to derive the MBQC description of arbitrary QAOA$_p$ we consider %
a single edge $(u, v)$ in the $k$-th layer, $k\in\{1,\dots,p\}$, with parameters $\gamma_m,\beta_m$
\begin{align*}
    &
    \vcenter{\hbox{%
        \externalizezx{qaoambqcdiagram01}{%
            %
        }
    }} \, .
\end{align*}
This is almost a graph state with single qubit operations, except for the two spiders 
$
\zx{\zxN{}\rar&[0.5em]\zxX{m'_u \pi}\rar&\zxZ{m_u\pi}\rar&[0.5em] \zxN{}}
$
on the top right.
In order to achieve the desired graph state with just single qubit rotations, we need to move these two spiders to the next layer.
For the benefit of the reader, we visually mark them with blue and red dots below.
\begin{align*}
    &
    \vcenter{\hbox{%
        \externalizezx{qaoambqcdiagram03}{%
            %
        }
    }} \; ,
\end{align*}
Note, that the measurement variables of the preceding ($k-1$)-th layer are denoted by $n$ whereas in the $k$-th layer they are denoted by $m$.
Hence, by moving the blue and red marked spiders to the next layer, they enter the diagrams above from the left $(k-1)$-th layer as $n'_u$, $n'_v$, $n_u$ and $n_v$.

We need to be careful about the 
$
\zx{\zxN{}\rar&[0.5em]\zxX{n'_u \pi}\rar&[0.5em] \zxN{}}
$
at the top of the above diagram, since they are connecting to all nodes in the neighborhood of $u$ without $v$, $\mathcal{N}(u) \setminus v$. 
Fortunately, these spiders stay local in the following sense. 
Consider one of the legs with an 
$
\zx{\zxN{}\rar&[0.5em]\zxX{n'_u \pi}\rar&[0.5em] \zxN{}}
$
leaving the above diagram at the top, say to a spider stemming from the node $w \in \mathcal{N}(u) \setminus v$.
There will be a similar spider
$
\zx{\zxN{}\rar&[0.5em]\zxX{n'_w \pi}\rar&[0.5em] \zxN{}}
$
connected via an Hadamard node
\begin{align}
    \begin{ZX}[ampersand replacement=\&]
        \zxN{}\rar        \&[0.5em] 
        \zxX{n'_u \pi}\rar \&[0.5em] 
        \zxH{} \rar       \&[0.5em]
        \zxX{n'_w \pi}\rar \&[0.5em] 
        \zxN{}
    \end{ZX}
    &
    \stackrel{\hadamardrule}{=}%
    \begin{ZX}[ampersand replacement=\&]
        \zxN{}\rar        \&[0.5em] 
        \zxX{n'_u \pi}\rar \&[0.5em] 
        \zxZ{n'_w \pi}\rar \&[0.5em] 
        \zxH{} \rar       \&[0.5em]
        \zxN{}
    \end{ZX}
    \\[2ex]
    &
    \stackrel{\picopyrule}{=}%
    \begin{ZX}[ampersand replacement=\&]
        \zxN{}\rar         \&[0.5em] 
        \zxZ{(-1)^{n'_u} n'_w \pi}\rar \&[0.5em] 
        \zxX{n'_u \pi}\rar  \&[0.5em] 
        \zxH{} \rar        \&[0.5em]
        \zxN{}
    \end{ZX}
    \\
    &=
    \begin{ZX}[ampersand replacement=\&]
        \zxN{}\rar         \&[0.5em] 
        \zxZ{n'_w \pi}\rar \&[0.5em] 
        \zxH{} \rar        \&[0.5em]
        \zxZ{n'_u \pi}\rar  \&[0.5em] 
        \zxN{}
    \end{ZX} \;\; . \label{eqn:edge_to_environment_identiy}
\end{align}
Where in the last step, we have used 
\begin{equation}
    \begin{ZX}[ampersand replacement=\&]
        \zxN{}\rar         \&
        \zxZ{-\pi}    \rar \&
        \zxN{}
    \end{ZX}
    \stackrel{\substack{\spiderrule\\ \idrule}}{=}%
    \begin{ZX}[ampersand replacement=\&]
        \zxN{}\rar         \&
        \zxZ{-\pi}    \rar \&
        \zxZ{2\pi}    \rar \&
        \zxN{}
    \end{ZX}
    \stackrel{\spiderrule}{=}%
    \begin{ZX}[ampersand replacement=\&]
        \zxN{}\rar         \&
        \zxZ{\pi}    \rar \&
        \zxN{}
    \end{ZX} \;\; .
\end{equation}
With this, we can write
\begin{align*}
    \vcenter{\hbox{%
        \externalizezx{qaoambqcdiagram06}{%
            \begin{tikzpicture}[node distance=8.0ex and 8.0em]%
                \def\braceraise{7ex}
\def\bracelabelraise{8ex}
    \node[on grid, name=u-ps, zident];
    \node[on grid, name=uv-ps, below=of u-ps, zident];
    \node[on grid, name=v-ps, below=of uv-ps, zident];
    \draw (u-ps) -- (uv-ps) node[midway, hadamard];
    \draw (uv-ps) -- (v-ps) node[midway, hadamard];

    \draw[gray](u-ps) -- ++ (120:4ex) node[midway, hadamard, draw=gray];
    \draw[gray](u-ps) -- ++ (60:4ex) node[midway, hadamard, draw=gray];
    \node[gray, above=2ex of u-ps] {\footnotesize $\dots$};
    \draw[gray](v-ps) -- ++ (-120:4ex) node[midway, hadamard, draw=gray];
    \draw[gray](v-ps) -- ++ (-60:4ex) node[midway, hadamard, draw=gray];
    \node[gray, below=2ex of v-ps] {\footnotesize $\dots$};

    \node[name=uv-psr, right=2em of uv-ps, zspider] {$n'_u + n'_v$};
    \node[name=uv-psrr, right=1.5em of uv-psr, xspider] {$\gamma_m$};
    \node[name=uv-psrrr, right=2em of uv-psrr, xspider] {$m_{uv}\pi$};
    \draw(uv-ps) -- (uv-psr);
    \draw(uv-psr) -- (uv-psrr);
    \draw(uv-psrr) -- (uv-psrrr);

    \node[name=u-psl, left=0.1em of u-ps, empty];
    \node[name=v-psl, left=0.1em of v-ps, empty];
    \node[name=u-psl, left=0.5em of u-psl]{\footnotesize$u$};
    \node[name=v-psl, left=0.5em of v-psl]{\footnotesize$v$};
    \draw(u-ps) -- (u-psl);
    \draw(v-ps) -- (v-psl);

    \node[on grid, name=u1-m, right=of u-ps, zident];
    \draw(u-ps) -- (u1-m) node[midway, hadamard];

    \node[name=u-m, xshift=1ex, below right=5ex of u-ps, zspider] {$\mathcal{P}_u \pi$};
    \node[name=u-mr, right=4ex of u-m, xspider] {$n'_u \pi$};
    \node[name=u-mrr, right=4ex of u-mr, zspider] {$n_u \pi$};
    \node[name=u-mrrr, right=4ex of u-mrr, zspider] {$m_{uv} \pi$};
    \node[name=u-mrrrr, right=4ex of u-mrrr, zspider] {$m_u \pi$};
    \node[name=u1-mr, above=2ex of u1-m, zspider] {$n'_u$};
    \node[name=u1-mrr, above=2ex of u1-mr, zspider] {$(-1)^{m_u} \beta_m$};
    \node[name=u1-mrrr, above=2ex of u1-mrr, zspider] {$m'_u \pi$};
    \draw(u-ps) -- (u-m);
    \draw(u-m) -- (u-mr);
    \draw(u-mr) -- (u-mrr);
    \draw(u-mrr) -- (u-mrrr);
    \draw(u-mrrr) -- (u-mrrrr);
    \draw(u1-m) -- (u1-mr);
    \draw(u1-mr) -- (u1-mrr);
    \draw(u1-mrr) -- (u1-mrrr);

    \node[on grid, name=v1-m, right=of v-ps, empty];
    \node[on grid, name=v2-m, right=of v1-m, empty];
    \node[on grid, name=v2-mr, right=of v2-m, empty];
    \node[on grid, name=v2-mrr, right=of v2-mr, empty];
    \draw[dashed](v-ps) -- ++(2em, 0ex);
    \node[right=2.0em of v-ps, anchor=west, gray]{\footnotesize  same as above for $u \leftrightarrow v$};

    \node[on grid, name=u-psn, right=of u1-m, zident, draw=gray];
    \node[on grid, name=uv-psn, below=of u-psn,  zident, draw=gray];
    \node[on grid, name=v-psn, below=of uv-psn,  zident, draw=gray];
    \draw[gray] (u-psn) -- (uv-psn) node[midway, hadamard, draw=gray];
    \draw[gray] (uv-psn) -- (v-psn) node[midway, hadamard, draw=gray];
    \draw[gray, dashed] (u-psn) -- ++(0:2ex);
    \draw[gray, dashed] (v-psn) -- ++(0:2ex);

    \draw(u1-m) -- (u-psn) node[midway, hadamard];
    \draw[dashed](v-psn) -- ++(180:4ex);

    \draw[gray](u-psn) -- ++ (120:4ex) node[midway, hadamard, draw=gray];
    \draw[gray](u-psn) -- ++ (60:4ex) node[midway, hadamard, draw=gray];
    \node[gray, above=2ex of u-psn] {\footnotesize $\dots$};
    \draw[gray](v-psn) -- ++ (-120:4ex) node[midway, hadamard, draw=gray];
    \draw[gray](v-psn) -- ++ (-60:4ex) node[midway, hadamard, draw=gray];
    \node[gray, below=2ex of v-psn] {\footnotesize $\dots$};

    \def\braceraise{7ex}
    \def\bracelabelraise{8ex}
    \draw[decorate,decoration={brace,mirror,raise=\braceraise},thick]%
         ([xshift=-1em]v-ps.center) -- ([xshift=12em]v-ps.center)%
         node[midway, below=\bracelabelraise, anchor=north] {\footnotesize $k$-th QAOA layer};
    \draw[decorate,decoration={brace,mirror,raise=\braceraise},thick, gray]%
         ([xshift=-1.5em]v-psn.center) -- ([xshift=2em]v-psn.center)%
         node[midway, below=\bracelabelraise, anchor=north] {\footnotesize  \color{gray} next ps layer};

            \end{tikzpicture}
        }
    }}
\end{align*}
where the parity of all the contribution from the neighboring edges is given by
\begin{equation}
    \mathcal{P}_u := \sum_{w\in \mathcal{N}_u \setminus v} n'_w \,.
\end{equation}
Note, that a deterministic measurement pattern emerges by the following order of measurement, indicated by the corresponding binary variables
\begin{equation}
    \dots,%
    \underbrace{n'_{uv}, n_u, n_v, n'_u, n'_v}_{(k-1)\text{-th layer}},%
    \underbrace{m'_{uv}, m_u, m_v, m'_u, m'_v}_{(k)\text{-th layer}}%
    ,\dots \quad .
\end{equation}
A similar derivation (see Appendix~\ref{sec:proof_qaoa_mbqc_qubo}) for the general QUBO case leads to
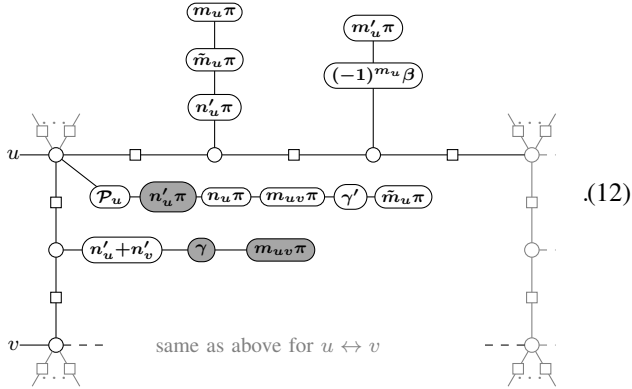
\begin{align}
    \vcenter{\hbox{%
        \externalizezx{qaoambqcqubo06}{%
            \begin{tikzpicture}[node distance=8.0ex and 6.0em]%
                    \node[name=u-ps, zident];
    \node[on grid, name=uv-ps, below=of u-ps, zident];
    \node[on grid, name=v-ps, below=of uv-ps, zident];
    \draw (u-ps) -- (uv-ps) node[midway, hadamard];
    \draw (uv-ps) -- (v-ps) node[midway, hadamard];

    \draw[gray](u-ps) -- ++ (120:4ex) node[midway, hadamard, draw=gray];
    \draw[gray](u-ps) -- ++ (60:4ex) node[midway, hadamard, draw=gray];
    \node[gray, above=2ex of u-ps] {\footnotesize $\dots$};
    \draw[gray](v-ps) -- ++ (-120:4ex) node[midway, hadamard, draw=gray];
    \draw[gray](v-ps) -- ++ (-60:4ex) node[midway, hadamard, draw=gray];
    \node[gray, below=2ex of v-ps] {\footnotesize $\dots$};

    \node[on grid, name=uv-psr, right=2.5em of uv-ps, zspider] {$n'_u + n'_v$};
    \node[on grid, name=uv-psrr, right=3em of uv-psr, xspider] {$\gamma$};
    \node[on grid, name=uv-psrrr, right=3em of uv-psrr, xspider] {$m_{uv}\pi$};
    \draw(uv-ps) -- (uv-psr);
    \draw(uv-psr) -- (uv-psrr);
    \draw(uv-psrr) -- (uv-psrrr);
    \node[name=u-psl, left=1.1em of u-ps]{\footnotesize$u$};
    \node[name=v-psl, left=1.1em of v-ps]{\footnotesize$v$};
    \draw(u-ps) -- (u-psl);
    \draw(v-ps) -- (v-psl);

    \node[on grid, name=u2-sq, right=of u-ps, zident];
    \draw(u-ps) -- (u2-sq) node[midway, hadamard];

    \node[on grid, name=u-sq, xshift=1ex, below right=5ex of u-ps, zspider] {$\mathcal{P}_u$};
    \node[on grid, name=u-sqr, right=2.2em of u-sq, xspider] {$n'_u \pi$};
    \node[on grid, name=u-sqrr, right=2.2em of u-sqr, zspider] {$n_u \pi$};
    \node[on grid, name=u-sqrrr, right=2.5em of u-sqrr, zspider] {$m_{uv} \pi$};
    \node[on grid, name=u-sqrrrr, right=2.2em of u-sqrrr, zspider] {$\gamma'$};
    \node[on grid, name=u-sqrrrrr, right=2em of u-sqrrrr, zspider] {$\tilde{m}_u \pi$};
    \draw(u-sq) -- (u-ps);
    \draw(u-sqr) -- (u-sq);
    \draw(u-sqrr) -- (u-sqr);
    \draw(u-sqrrr) -- (u-sqrr);
    \draw(u-sqrrrr) -- (u-sqrrr);
    \draw(u-sqrrrrr) -- (u-sqrrrr);

    \node[on grid, name=u3-m, right=of u2-sq, zident];
    \draw(u2-sq) -- (u3-m) node[midway, hadamard];

    \node[on grid, name=u-m, above=4ex of u2-sq, zspider] {$n'_u \pi$};
    \node[on grid, name=u-mr, above=4ex of u-m, zspider] {$\tilde{m}_u \pi$};
    \node[on grid, name=u-mrr, above=4ex of u-mr, zspider] {$m_u \pi$};
    \node[on grid, name=u3-mr, above=3.0em of u3-m, zspider] {$(-1)^{m_u} \beta$};
    \node[on grid, name=u3-mrr, above=4ex of u3-mr, zspider] {$m'_u \pi$};
    \draw(u2-sq) -- (u-m);
    \draw(u-m) -- (u-mr);
    \draw(u-mr) -- (u-mrr);
    \draw(u3-m) -- (u3-mr);
    \draw(u3-mr) -- (u3-mrr);

    \draw[dashed](v-ps) -- ++(2em, 0ex);
    \node[right=3.5em of v-ps, anchor=west, gray]{\footnotesize  same as above for $u \leftrightarrow v$};

    \node[on grid, name=u-psn, right=of u3-m, zident, draw=gray];
    \node[on grid, name=uv-psn, below=of u-psn,  zident, draw=gray];
    \node[on grid, name=v-psn, below=of uv-psn,  zident, draw=gray];
    \draw[gray] (u-psn) -- (uv-psn) node[midway, hadamard, draw=gray];
    \draw[gray] (uv-psn) -- (v-psn) node[midway, hadamard, draw=gray];
    \draw[gray, dashed] (u-psn) -- ++(0:2ex);
    \draw[gray, dashed] (uv-psn) -- ++(0:2ex);
    \draw[gray, dashed] (v-psn) -- ++(0:2ex);

    \draw(u3-m) -- (u-psn) node[midway, hadamard];
    \draw[dashed](v-psn) -- ++(180:4ex);

    \draw[gray](u-psn) -- ++ (120:4ex) node[midway, hadamard, draw=gray];
    \draw[gray](u-psn) -- ++ (60:4ex) node[midway, hadamard, draw=gray];
    \node[gray, above=2ex of u-psn] {\footnotesize $\dots$};
    \draw[gray](v-psn) -- ++ (-120:4ex) node[midway, hadamard, draw=gray];
    \draw[gray](v-psn) -- ++ (-60:4ex) node[midway, hadamard, draw=gray];
    \node[gray, below=2ex of v-psn] {\footnotesize $\dots$};

            \end{tikzpicture}
        }
    }} \; .  \label{eqn:qaoa_mbqc_qubo}
\end{align}
With this, we have derived a general formula for QAOA with arbitrary number of layers in the MBQC paradigm.
Note, that the resource graph state we are using here is not a planar graph in general. It is directly derived from the interaction graph of the problem Hamiltonian.
However, it can be compiled in a straight-forward way into planar graphs of the target hardware via un-fusing nodes~\cite{zilk2022compiler}.

\subsection{Resource Estimates}
Let us now consider the resource requirements of our protocol, again neglecting any single $Z$ terms for the moment.  
For each QAOA layer, we introduced one 
additional ancilla qubit for each edge in the interaction graph and another 
two for each vertex.
Hence, the number of qubits needed is at most
\begin{equation}
    \mathcal{N}_\mathrm{Q} \leq p (|E| + 2|V|) \; ,
    \label{eq:qaoa_mbqc_number_qubits} 
\end{equation}
where $p$ is the number of QAOA layers. Note that this bound is conservative in that no reuse of ancilla qubits is assumed.
Similarly, for each QAOA layer, we have two entangling CZ gates for each edge and two for each vertex in the interaction graph. 
Hence the total number of entangling CZ gates, i.e.~ the number of edges in the graph state %
is at most 
\begin{equation}
    \mathcal{N}_\mathrm{E} \leq p (2|E| + 2 |V|). %
\label{eq:qaoa_mbqc_number_entangling_gates} 
\end{equation}
For the general QUBO case that includes single-qubit rotations in the phase operator, %
we %
require 
at most 
one additional qubit and entangling gate for each vertex %
per QAOA layer.

In comparison, quantum gate model implementations of QAOA 
require $|V|$ logical qubits, and at least $2p|E|$ entangling gates for standard compilations~\cite{hadfield2018quantum}, though we emphasize that compilation considerations related to limited hardware connectivity, fault tolerance, and gate set can add considerable additional overhead to these estimates. Hence as expected the gate-model approach requires fewer circuit resources. 
However, the usual considerations for potentially attractive aspects of MBQC still apply. For example, the resource state can be %
finely-tuned and deployed for a large variety of problems as long as the original problem graph is a minor of the resource state graph. 
Also, the number of qubits required can be significantly reduced in some cases by reusing qubits after measurement~\cite{decross2023qubit}. 
Moreover, our approach can be used as a starting point for further investigations into mixed, potentially more sophisticated paradigms that lay between the gate-model and MBQC, with more explicit resource tradeoffs,
as well as algorithm adaptations further tailored to specific hardware platforms.

\section{QAOA for MIS in the MBQC paradigm}
\label{sec:qaoa_mbqc_mis}
In this section we  discuss how to incorporate the quantum alternating operator ansatz~\cite{hadfield2019quantum} into the measurement based quantum computing paradigm, for a particular problem with hard constraints, the maximum independent set (MIS) problem.
In this generalization of QAOA, mixers can be constructed as ordered products of partial mixers that each act on a subset of qubits, and are designed as to only induce transitions between feasible (constraint-satisfying) states. 

In the MIS problem we are given a graph $G=(V, E)$ and we seek to find a large as possible subset of the vertices with no edges shared between them.
Following the construction of~\cite[Sec. 4.2.1]{hadfield2019quantum}, the partial mixing operator for each node $v\in V$ is given by the controlled unitary
\begin{equation}
    U_v (\beta) = \Lambda_{\mathcal{N}(v)} (\ee^{\ii\beta X_v}),
\end{equation}
where $\mathcal{N}(v)$ is the neighborhood of nodes adjacent to $v$ in the graph, and $\Lambda_{\mathcal{N}(v)}(\cdot)$ indicates that the $X$-rotation within is controlled by all of the bits in $\mathcal{N}(v)$ being set to $0$.
It can be shown using ZH-calculus~\cite{backens2018zh}, 
a closely related variant of ZX-calculus, 
that 
this partial mixing operator can be %
expressed as 
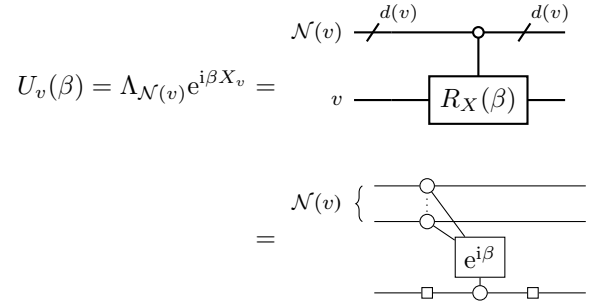
\begin{align*}
    U_v (\beta) = \Lambda_{\mathcal{N}(v)} \ee^{\ii\beta X_v}
    &=
    \vcenter{\hbox{%
            \begin{tikzcd}[ampersand replacement=\&]
    \lstick{\footnotesize $\mathcal{N}(v)$} \& \qwbundle{d(v)} \& \octrl{1}          \& \qwbundle{d(v)} \\
    \lstick{\footnotesize $v$}              \& \qw             \& \gate{R_X(\beta)}  \& \qw       \\
\end{tikzcd}

    }}
    \\
    &=
    \vcenter{\hbox{%
            \begin{tikzpicture}[node distance=3ex and 2.0em]%
    \node[name=l-nt, empty];
    \node[on grid, name=l-nb, below=of l-nt, empty];
    \node[on grid, name=ml-nt, right=of l-nt, zident];
    \node[on grid, name=ml-nb, right=of l-nb, zident];
    \node[on grid, name=mr-nt, right=of ml-nt, empty];
    \draw(l-nt) -- (ml-nt);
    \draw(l-nb) -- (ml-nb);
    \node[on grid, name=mr-nb, right=of ml-nb, empty];
    \node[on grid, name=r-nt, right=of mr-nt, empty];
    \node[on grid, name=r-nb, right=of mr-nb, empty];
    \node[on grid, name=rr-nt, right=of r-nt, empty];
    \node[on grid, name=rr-nb, right=of r-nb, empty];
    \draw(ml-nt) -- (rr-nt);
    \draw(ml-nb) -- (rr-nb);
    \node[on grid, name=ctrl, below=of mr-nb, rectangle, draw=black] {$\ee^{\ii \beta}$};
    \draw(ml-nt) -- (ctrl);
    \draw(ml-nb) -- (ctrl);
    \node[on grid, name=mr, below=of ctrl, zident];
    \draw(ctrl) -- (mr);
    \node[on grid, name=ml, left=of mr, hadamard];
    \draw(mr) -- (ml);
    \node[on grid, name=l, left=of ml, empty];
    \draw(l) -- (ml);
    \node[on grid, name=r, right=of mr, hadamard];
    \draw(r) -- (mr);
    \node[on grid, name=rr, right=of r, empty];
    \draw(rr) -- (r);

    \node[rotate=90, anchor=center, scale=0.5] at ($(ml-nt)!0.5!(ml-nb)$) {$\dots$};
    \draw[decorate, decoration={brace,mirror,raise=1ex}] (l-nt) -- (l-nb)%
         node[midway, left=2ex, anchor=east, rotate=00] {\footnotesize $\mathcal{N}(v$)};
\end{tikzpicture}

    }}
\end{align*}
Hence, similarly to the QUBO case,
this constitute the %
most important step toward the formulation of a quantum alternating operator ansatz for MIS in the MBQC paradigm. 
Here, the QAOA phase operator is comprised of single-qubit rotation gates and so can be constructed 
using the same techniques of Section~\ref{sec:qaoa_mbqc}. 
For the initial state, one generally desires a superposition of feasible states. 
For MIS one could use the product state corresponding to a classically determined approximate solution, i.e., an independent set, followed by an initial application of the mixer operator~\cite{hadfield2019quantum}. 
The overall QAOA circuit is then constructed analogously combining the phase and mixing operator layers in alternation. %

\section{Measurement-based Quantum Optimization}
\label{sec:outlook}
A wide variety of optimization problems can be mapped to QUBO problems~\cite{lucas2014ising,hadfield2021representation}. This includes numerous constrained optimization problems, which are typically mapped to QUBOs through the inclusion of additional penalty terms, and in some cases additional variables, among other techniques. These problems can then be tackled with QAOA in the MBQC paradigm applying the results of Section~\ref{sec:qaoa_mbqc}. 

For many constrained problems it is desirable to utilize more general QAOA circuits with mixers that preserve problem hard constraints~\cite{hadfield2019quantum}. The approach of Section~\ref{sec:qaoa_mbqc_mis} gives an initial prototype for similarly extending such constructions to the MBQC paradigm. 
A particular example are a number of optimization problems related to graph colorings, for which QAOA constructions often employ %
\textit{XY partial mixers}~\cite{hadfield2019quantum,wang2020x} given by $U_{uv}(\beta)=\ee^{\ii\beta (X_uX_v +Y_uY_v)}=\ee^{\ii\beta X_uX_v} \ee^{\ii\beta Y_uY_v}$. The operators $\ee^{\ii\beta X_uX_v}$ and $\ee^{\ii\beta Y_uY_v}$ can be derived and implemented in a measurement-based paradigm in particular by adapting the results for the $\ee^{\ii\beta Z_uZ_v}$ operators of Section~\ref{sec:qaoa_mbqc}.

Alternatively, one might also consider wider varieties of parameterized quantum circuits beyond QAOA, such as so called hardware-efficent ans\"atze~\cite{moll2018quantum,larose2022mixer,Dupont_2023QuantumCombOpt,maciejewski2023design}. Examples of hardware-efficent ans\"atze were previously analyzed with ZX-calculus in~\cite{stollenwerk2022diagrammatic}. 
While these ans\"atze are built from gate-model basic primitives, and one may proceed similarly in translating to MBQC, an important direction of future research is to explore ans\"atze which are analgously built from primitives that are natural in some sense with respect to MBQC platforms. 
Furthermore, it is worthwhile to continue to explore to what extent diagrammatic approaches to MBQC can prove fruitful for applications beyond combinatorial optimization such as machine learning~\cite{yeung2020diagrammatic,toumi2021diagrammatic,koch2022quantum,mantilla2023measurement,majumder2023variational} or quantum chemistry~\cite{yeung2020diagrammatic,toumi2021diagrammatic,koch2022quantum,mantilla2023measurement,majumder2023variational}.

Finally, several recent works consider so-called \emph{iterative quantum optimization}~\cite{bravyi2020obstacles,Dupont_2023QuantumCombOpt,brady2023iterative} wherein the quantum device is used to estimate a set of observable expectation values, rather than solve the problem directly per se. The expectation values are used to select a reduction step to apply, which results in a smaller problem, and the process iterated until the residual problem is small enough to be solved exactly. The expectation values are typically estimated by repeated preparation and measurement using the quantum device, which in principle can be %
obtained using 
a quantum circuit such as QAOA or other solvers such as quantum annealers or MBQC approaches~\cite{brady2023iterative}.

\section{Conclusion}
\label{sec:conclusion}
We investigated measurement-based approaches to quantum approximate optimization by focusing on %
methodically translating quantum circuit ans\"atze to MBQC patterns using ZX-calculus. 
Our main result obtained is a general formulation of QAOA %
as an MBQC protocol for arbitrary number of layers or algorithm parameters. While we showed the QUBO case explicitly as explained our techniques extend to higher-order cost functions. 
In addition, we outline how to extend our results to the quantum alternating operator ansatz for maximum independent set and further classes of constrained optimization problems. 
In particular we derived the partial mixing operator for %
MIS with the ZH-calculus, which constitutes the first step for a complete MBQC formulation analogously to the QUBO case. 
Finally, a motivating question for this work is whether one can directly design measurement-based protocols for optimization tailored to this setting that yield significant performance or resource advantages as compared to translations of quantum gate model algorithms; we are optimistic ZX-calculus and the techniques employed here will prove useful in illuminating this.

\section*{Acknowledgments}
SH is grateful for support from the NASA Ames Research Center, from NASA Academic Mission Services (NAMS) under Contract No. NNA16BD14C, and from the DARPA ONISQ program under interagency agreement IAA 8839, Annex 114.

\bibliographystyle{IEEEtran}
\bibliography{IEEEabrv,references}

\appendix
\section{Appendix}
Here we provide diagrammatic proofs of several identities given in the main text.
\subsection{Measurement based quantum computing example in ZX-Calculus}%
\label{sec:proof_mbqc_example}
We derive the MBQC example from above. Measuring qubit $i$ in Pauli basis $P\in\{X, Y, Z\}$ and storing the result in the binary variable $n\in\{0, 1\}$, is denoted by $\mathcal{M}^i_P\to n$
$\Lambda^i_n(U)$
denotes the application of a single qubit gate $U$ onto qubit $i$ if and only if the binary variable $n=1$.
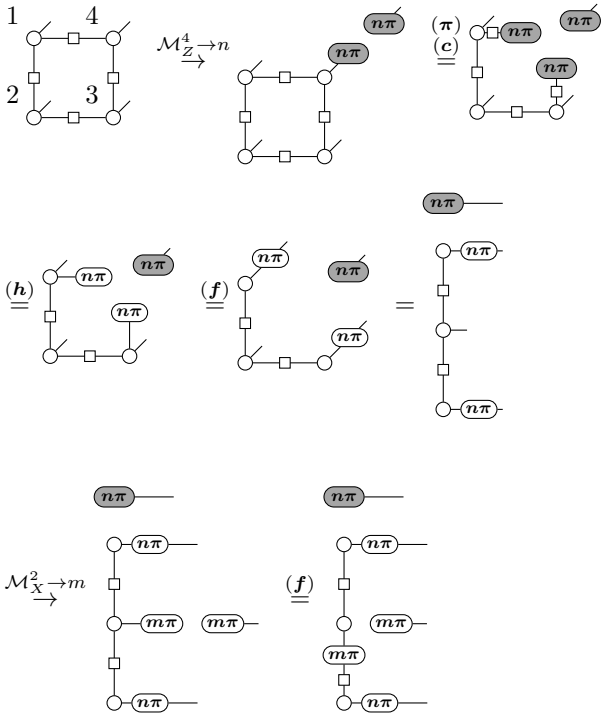
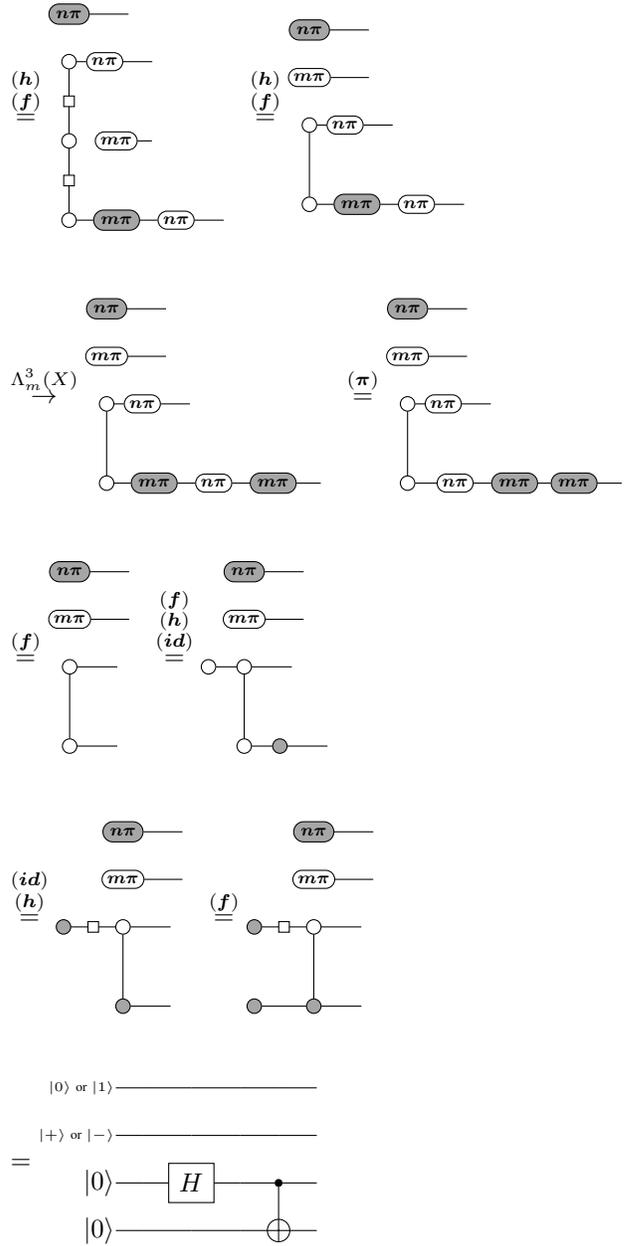
\begin{align*}
    &
    \vcenter{\hbox{%
        \externalize{appendixmbqcexamplediagram02}{%
            \begin{tikzpicture}[node distance=3.0em and 3.0em]%
                \node[on grid, name=1, zident, label=135:$1$];
\node[on grid, name=2, below=of 1, zident, label=135:$2$];
\node[on grid, name=3, right=of 2, zident, label=135:$3$];
\node[on grid, name=4, above=of 3, zident, label=135:$4$];

\draw (1) -- (2) node[midway, hadamard];
\draw (2) -- (3) node[midway, hadamard];
\draw (3) -- (4) node[midway, hadamard];
\draw (4) -- (1) node[midway, hadamard];

\draw (1) -- ++(45:2ex);
\draw (2) -- ++(45:2ex);
\draw (3) -- ++(45:2ex);
\draw (4) -- ++(45:2ex);

            \end{tikzpicture}
        }
    }} \notag
    \stackrel{\mathcal{M}^4_Z\to n}{\to}%
    \vcenter{\hbox{%
        \externalize{appendixmbqcexamplediagram03}{%
            \begin{tikzpicture}[node distance=3.0em and 3.0em]%
                \node[on grid, name=1, zident];
\node[on grid, name=2, below=of 1, zident];
\node[on grid, name=3, right=of 2, zident];
\node[on grid, name=4, above=of 3, zident];

\draw (1) -- (2) node[midway, hadamard];
\draw (2) -- (3) node[midway, hadamard];
\draw (3) -- (4) node[midway, hadamard];
\draw (4) -- (1) node[midway, hadamard];

\node[on grid, name=m4, xshift=2ex, above=2ex of 4, xspider] {$n \pi$};
\draw (1) -- ++(45:2ex);
\draw (2) -- ++(45:2ex);
\draw (3) -- ++(45:2ex);
\draw (4) -- (m4);

\node[on grid, name=m4c, xshift=3ex, above=2.5ex of m4, xspider] {$n \pi$};
\draw (m4c) -- ++(45:2ex);

            \end{tikzpicture}
        }
        \vspace*{-4ex}
    }} \notag
    \stackrel{\substack{\picopyrule \\ \copyrule}}{=}%
    \vcenter{\hbox{%
        \externalize{appendixmbqcexamplediagram04}{%
            \begin{tikzpicture}[node distance=3.0em and 3.0em]%
                \node[on grid, name=1, zident];
\node[on grid, name=2, below=of 1, zident];
\node[on grid, name=3, right=of 2, zident];
\node[on grid, name=4, above=of 3, empty];
\node[on grid, name=4l, above=of 3, xshift=-3ex, yshift=0ex, xspider]{$n\pi$};
\node[on grid, name=4r, above=of 3, xshift=0ex, yshift=-3ex, xspider]{$n\pi$};

\draw (1) -- (2) node[midway, hadamard];
\draw (2) -- (3) node[midway, hadamard];
\draw (3) -- (4r) node[midway, hadamard];
\draw (4l) -- (1) node[midway, hadamard];

\draw (1) -- ++(45:2ex);
\draw (2) -- ++(45:2ex);
\draw (3) -- ++(45:2ex);

\node[on grid, name=m4c, xshift=2ex, above=1.0ex of 4, xspider] {$n \pi$};
\draw (m4c) -- ++(45:2ex);

            \end{tikzpicture}
        }
    }} \notag
    \\[5ex]
    &%
    \stackrel{\hadamardrule}{=}%
    \vcenter{\hbox{%
        \externalize{appendixmbqcexamplediagram05}{%
            \begin{tikzpicture}[node distance=3.0em and 3.0em]%
                \node[on grid, name=1, zident];
\node[on grid, name=2, below=of 1, zident];
\node[on grid, name=3, right=of 2, zident];
\node[on grid, name=4, above=of 3, empty];
\node[on grid, name=4l, above=of 3, xshift=-3ex, yshift=0ex, zspider]{$n\pi$};
\node[on grid, name=4r, above=of 3, xshift=0ex, yshift=-3ex, zspider]{$n\pi$};

\draw (1) -- (2) node[midway, hadamard];
\draw (2) -- (3) node[midway, hadamard];
\draw (3) -- (4r);
\draw (4l) -- (1);

\draw (1) -- ++(45:2ex);
\draw (2) -- ++(45:2ex);
\draw (3) -- ++(45:2ex);

\node[on grid, name=m4c, xshift=2ex, above=1.0ex of 4, xspider] {$n \pi$};
\draw (m4c) -- ++(45:2ex);

            \end{tikzpicture}
        }
    }} \notag
    \stackrel{\spiderrule}{=}%
    \vcenter{\hbox{%
        \externalize{appendixmbqcexamplediagram06}{%
            \begin{tikzpicture}[node distance=3.0em and 3.0em]%
                \node[on grid, name=1, zident];
\node[on grid, name=2, below=of 1, zident];
\node[on grid, name=3, right=of 2, zident];
\node[on grid, name=4, above=of 3, empty];

\draw (1) -- (2) node[midway, hadamard];
\draw (2) -- (3) node[midway, hadamard];
\draw (2) -- ++(45:2ex);

\node[on grid, name=1r, above right=3ex of 1, zspider]{$n\pi$};
\node[on grid, name=3r, above right=3ex of 3, zspider]{$n\pi$};
\draw (1) -- (1r);
\draw (1r) -- ++(45:2ex);
\draw (3) -- (3r);
\draw (3r) -- ++(45:2ex);

\node[on grid, name=m4c, xshift=2ex, above=1.0ex of 4, xspider] {$n \pi$};
\draw (m4c) -- ++(45:2ex);

            \end{tikzpicture}
        }
    }} \notag
    =%
    \vcenter{\hbox{%
        \externalize{appendixmbqcexamplediagram07}{%
            \begin{tikzpicture}[node distance=3.0em and 3.0em]%
                \node[on grid, name=1, zident];
\node[on grid, name=2, below=of 1, zident];
\node[on grid, name=3, below=of 2, zident];

\draw (1) -- (2) node[midway, hadamard];
\draw (2) -- (3) node[midway, hadamard];
\draw (2) -- ++(0:2ex);

\node[on grid, name=1r, right=3ex of 1, zspider]{$n\pi$};
\node[on grid, name=3r, right=3ex of 3, zspider]{$n\pi$};
\draw (1) -- (1r);
\draw (1r) -- ++(0:2ex);
\draw (3) -- (3r);
\draw (3r) -- ++(0:2ex);

\node[on grid, name=4, above=4ex of 1, xspider] {$n \pi$};
\draw (4) -- ++(0:5ex);

            \end{tikzpicture}
        }
    }} \notag
    \\[5ex]
    &
    \stackrel{\mathcal{M}^2_X\to m}{\to}%
    \vcenter{\hbox{%
        \externalize{appendixmbqcexamplediagram08}{%
            \begin{tikzpicture}[node distance=3.0em and 3.0em]%
                \node[on grid, name=1, zident];
\node[on grid, name=2, below=of 1, zident];
\node[on grid, name=3, below=of 2, zident];

\draw (1) -- (2) node[midway, hadamard];
\draw (2) -- (3) node[midway, hadamard];
\node[on grid, name=m2, right=4ex of 2, zspider] {$m \pi$};
\draw (2) -- (m2);
\node[name=m2c, right=1.5em of m2, zspider] {$m \pi$};
\draw (m2c) -- ++(0:3ex);

\node[on grid, name=1r, right=3ex of 1, zspider]{$n\pi$};
\node[on grid, name=3r, right=3ex of 3, zspider]{$n\pi$};
\draw (1) -- (1r);
\draw (1r) -- ++(0:4ex);
\draw (3) -- (3r);
\draw (3r) -- ++(0:4ex);

\node[on grid, name=4, above=4ex of 1, xspider] {$n \pi$};
\draw (4) -- ++(0:5ex);

            \end{tikzpicture}
        }
    }} \notag
    \stackrel{\spiderrule}{=}%
    \vcenter{\hbox{%
        \externalize{appendixmbqcexamplediagram09}{%
            \begin{tikzpicture}[node distance=3.0em and 3.0em]%
                \node[on grid, name=1, zident];
\node[on grid, name=2, below=of 1, zident];
\node[on grid, name=3, below=of 2, zident];

\draw (1) -- (2) node[midway, hadamard];
\node[name=m2, below=2ex of 2, zspider] {$m \pi$};
\draw (2) -- (m2);
\draw (m2) -- (3) node[midway, hadamard];

\node[name=m2c, right=1.5em of 2, zspider] {$m \pi$};
\draw (m2c) -- ++(0:3ex);

\node[on grid, name=1r, right=3ex of 1, zspider]{$n\pi$};
\node[on grid, name=3r, right=3ex of 3, zspider]{$n\pi$};
\draw (1) -- (1r);
\draw (1r) -- ++(0:4ex);
\draw (3) -- (3r);
\draw (3r) -- ++(0:4ex);

\node[on grid, name=4, above=4ex of 1, xspider] {$n \pi$};
\draw (4) -- ++(0:5ex);

            \end{tikzpicture}
        }
    }} \notag
    \\[5ex]
    &%
    \stackrel{\substack{\hadamardrule \\ \spiderrule}}{=}%
    \vcenter{\hbox{%
        \externalize{appendixmbqcexamplediagram10}{%
            \begin{tikzpicture}[node distance=3.0em and 3.0em]%
                \node[on grid, name=1, zident];
\node[on grid, name=2, below=of 1, zident];
\node[on grid, name=3, below=of 2, zident];

\draw (1) -- (2) node[midway, hadamard];
\draw (2) -- (3) node[midway, hadamard];

\node[name=m2c, right=1.5em of 2, zspider] {$m \pi$};
\draw (m2c) -- ++(0:3ex);

\node[on grid, name=1r, right=3ex of 1, zspider]{$n\pi$};
\node[on grid, name=3r, right=4ex of 3, xspider] {$m \pi$};
\node[on grid, name=3rr, right=5ex of 3r, zspider]{$n\pi$};
\draw (1) -- (1r);
\draw (1r) -- ++(0:4ex);
\draw (3) -- (3r);
\draw (3r) -- (3rr);
\draw (3rr) -- ++(0:4ex);

\node[on grid, name=4, above=4ex of 1, xspider] {$n \pi$};
\draw (4) -- ++(0:5ex);

            \end{tikzpicture}
        }
    }} \notag
    \stackrel{\substack{\hadamardrule \\ \spiderrule}}{=}%
    \vcenter{\hbox{%
        \externalize{appendixmbqcexamplediagram11}{%
            \begin{tikzpicture}[node distance=3.0em and 3.0em]%
                \node[on grid, name=1, zident];
\node[on grid, name=3, below=of 1, zident];
\draw (1) -- (3);

\node[on grid, name=1r, right=3ex of 1, zspider]{$n\pi$};
\node[on grid, name=3r, right=4ex of 3, xspider] {$m \pi$};
\node[on grid, name=3rr, right=5ex of 3r, zspider]{$n\pi$};
\draw (1) -- (1r);
\draw (1r) -- ++(0:4ex);
\draw (3) -- (3r);
\draw (3r) -- (3rr);
\draw (3rr) -- ++(0:4ex);

\node[on grid, name=2, above=4ex of 1, zspider] {$m \pi$};
\draw (2) -- ++(0:5ex);
\node[on grid, name=4, above=4ex of 2, xspider] {$n \pi$};
\draw (4) -- ++(0:5ex);

            \end{tikzpicture}
        }
    }} \notag
    \\[5ex]
    &%
    \stackrel{\Lambda^3_m(X)}{\to}%
    \vcenter{\hbox{%
        \externalize{appendixmbqcexamplediagram12}{%
            \begin{tikzpicture}[node distance=3.0em and 3.0em]%
                \node[on grid, name=1, zident];
\node[on grid, name=3, below=of 1, zident];
\draw (1) -- (3);

\node[on grid, name=1r, right=3ex of 1, zspider]{$n\pi$};
\node[on grid, name=3r, right=4ex of 3, xspider] {$m\pi$};
\node[on grid, name=3rr, right=5ex of 3r, zspider]{$n\pi$};
\node[on grid, name=3rrr, right=5ex of 3rr, xspider]{$m\pi$};
\draw (1) -- (1r);
\draw (1r) -- ++(0:4ex);
\draw (3) -- (3r);
\draw (3r) -- (3rr);
\draw (3rr) -- (3rrr);
\draw (3rrr) -- ++(0:4ex);

\node[on grid, name=2, above=4ex of 1, zspider] {$m\pi$};
\draw (2) -- ++(0:5ex);
\node[on grid, name=4, above=4ex of 2, xspider] {$n\pi$};
\draw (4) -- ++(0:5ex);

            \end{tikzpicture}
        }
    }} \notag
    \stackrel{\picopyrule}{=}%
    \vcenter{\hbox{%
        \externalize{appendixmbqcexamplediagram13}{%
            \begin{tikzpicture}[node distance=3.0em and 3.0em]%
                \node[on grid, name=1, zident];
\node[on grid, name=3, below=of 1, zident];
\draw (1) -- (3);

\node[on grid, name=1r, right=3ex of 1, zspider]{$n\pi$};
\node[on grid, name=3r, right=4ex of 3, zspider] {$n\pi$};
\node[on grid, name=3rr, right=5ex of 3r, xspider]{$m\pi$};
\node[on grid, name=3rrr, right=5ex of 3rr, xspider]{$m\pi$};
\draw (1) -- (1r);
\draw (1r) -- ++(0:4ex);
\draw (3) -- (3r);
\draw (3r) -- (3rr);
\draw (3rr) -- (3rrr);
\draw (3rrr) -- ++(0:4ex);

\node[on grid, name=2, above=4ex of 1, zspider] {$m\pi$};
\draw (2) -- ++(0:5ex);
\node[on grid, name=4, above=4ex of 2, xspider] {$n\pi$};
\draw (4) -- ++(0:5ex);

            \end{tikzpicture}
        }
    }} \notag
    \\[5ex]
    &%
    \stackrel{\spiderrule}{=}%
    \vcenter{\hbox{%
        \externalize{appendixmbqcexamplediagram14}{%
            \begin{tikzpicture}[node distance=3.0em and 3.0em]%
                \node[on grid, name=1, zident];
\node[on grid, name=3, below=of 1, zident];
\draw (1) -- (3);

\draw (1) -- ++(0:4ex);
\draw (3) -- ++(0:4ex);

\node[on grid, name=2, above=4ex of 1, zspider] {$m\pi$};
\draw (2) -- ++(0:5ex);
\node[on grid, name=4, above=4ex of 2, xspider] {$n\pi$};
\draw (4) -- ++(0:5ex);

            \end{tikzpicture}
        }
    }} \notag
    \stackrel{\substack{\spiderrule \\ \hadamardrule \\ \idrule}}{=}%
    \vcenter{\hbox{%
        \externalize{appendixmbqcexamplediagram15}{%
            \begin{tikzpicture}[node distance=3.0em and 3.0em]%
                \node[on grid, name=1, zident];
\node[on grid, name=3, below=of 1, zident];
\draw (1) -- (3);

\node[on grid, name=1l, left=3ex of 1, zident];
\node[on grid, name=3r, right=3ex of 3, xident];
\draw (1) -- (1l);
\draw (3) -- (3r);
\draw (1) -- ++(0:4ex);
\draw (3r) -- ++(0:4ex);

\node[on grid, name=2, above=4ex of 1, zspider] {$m\pi$};
\draw (2) -- ++(0:5ex);
\node[on grid, name=4, above=4ex of 2, xspider] {$n\pi$};
\draw (4) -- ++(0:5ex);

            \end{tikzpicture}
        }
    }} \notag
    \\[5ex]
    &%
    \stackrel{\substack{\idrule \\ \hadamardrule}}{=}%
    \vcenter{\hbox{%
        \externalize{appendixmbqcexamplediagram16}{%
            \begin{tikzpicture}[node distance=3.0em and 3.0em]%
                \node[on grid, name=1, zident];
\node[on grid, name=3, below=of 1, xident];
\draw (1) -- (3);

\node[on grid, name=1l, left=5ex of 1, xident];
\draw (1) -- (1l) node[midway, hadamard];
\draw (1) -- ++(0:4ex);
\draw (3) -- ++(0:4ex);

\node[on grid, name=2, above=4ex of 1, zspider] {$m\pi$};
\draw (2) -- ++(0:5ex);
\node[on grid, name=4, above=4ex of 2, xspider] {$n\pi$};
\draw (4) -- ++(0:5ex);

            \end{tikzpicture}
        }
    }} \notag
    \stackrel{\spiderrule}{=}%
    \vcenter{\hbox{%
        \externalize{appendixmbqcexamplediagram17}{%
            \begin{tikzpicture}[node distance=3.0em and 3.0em]%
                \node[on grid, name=1, zident];
\node[on grid, name=3, below=of 1, xident];
\draw (1) -- (3);

\node[on grid, name=1l, left=5ex of 1, xident];
\node[on grid, name=3l, left=5ex of 3, xident];
\draw (1) -- (1l) node[midway, hadamard];
\draw (3) -- (3l);
\draw (1) -- ++(0:4ex);
\draw (3) -- ++(0:4ex);

\node[on grid, name=2, above=4ex of 1, zspider] {$m\pi$};
\draw (2) -- ++(0:5ex);
\node[on grid, name=4, above=4ex of 2, xspider] {$n\pi$};
\draw (4) -- ++(0:5ex);

            \end{tikzpicture}
        }
    }} \notag
    \\[5ex]
    &=%
    \vcenter{\hbox{%
        \externalizezx{appendixmbqcexamplecircuitfinal}{%
            \begin{tikzcd}[thin lines, row sep={4ex,between origins}, column sep=1.0em, ampersand replacement=\&]
    \lstick{\tiny \ket{0} \text{or} \ket{1}} \& \qw \& \qw       \& \qw \& \qw       \& \qw \&  \\
    \lstick{\tiny \ket{+} \text{or} \ket{-}} \& \qw \& \qw       \& \qw \& \qw       \& \qw \&  \\
    \lstick{\ket{0}}                         \& \qw \& \gate{H}  \& \qw \& \ctrl{1}  \& \qw \&  \\
    \lstick{\ket{0}}                         \& \qw \& \qw       \& \qw \& \targ{}   \& \qw \& 
\end{tikzcd}

        }
    }} \notag
\end{align*}
\subsection{Phase-separation operator in MBQC}
For the conversion of the phase-gadget into the MBQC paradignm, we show the derivation of \eqref{eqn:phase_gadget_zx} and \eqref{eqn:phase-gadget-to-mbqc}.
\subsubsection{Proof of \eqref{eqn:phase_gadget_zx}}%
\label{sec:proof_phase_gadget_zx}
\begin{align*}
    \vcenter{\hbox{%
        \externalizezx{appendixphasegadgetdiagram00}{%
            \begin{tikzcd}[thin lines, column sep=1em, row sep={4.0ex,between origins}, ampersand replacement=\&]
    \qw \& \ctrl{1} \& \qw \& \ctrl{1} \& \qw \\
    \qw \& \targ{}  \& \gate{R_Z(\gamma)} \& \targ{} \& \qw 
\end{tikzcd}
        }
    }}
    &=%
    \vcenter{\hbox{%
        \externalizezx{appendixphasegadgetdiagram01}{%
            \begin{ZX}[zx row sep=2.5ex, zx column sep=1em, ampersand replacement=\&]
    \zxN{} \rar \& \zxZ{} \dar \rar \& \zxN{}       \rar \& \zxZ{} \dar \rar \& \zxN{} \\
    \zxN{} \rar \& \zxX{}      \rar \& \zxZ{\gamma} \rar \& \zxX{}      \rar \& \zxN{} 
\end{ZX}
        }
    }}
    \\[2ex]
    \stackrel{\spiderrule}{=} %
    \vcenter{\hbox{%
        \externalizezx{appendixphasegadgetdiagram02}{%
            \begin{ZX}[zx row sep=2.5ex, zx column sep=1em, ampersand replacement=\&]
    \zxN{} \rar \& \zxN{} \rar            \& \zxZ{}       \rar \& \zxN{} \rar                 \& \zxN{} \\
    \zxN{} \rar \& \zxX{} \ar[ru] \ar[rd] \& \zxZ{\gamma} \dar \& \zxX{} \ar[lu] \ar[ld] \rar \& \zxN{} \\
                \&                        \& \zxZ{}            \&                             \&
\end{ZX}
        }
    }}
    & \stackrel{\spiderrule}{=} %
    \vcenter{\hbox{%
        \externalizezx{appendixphasegadgetdiagram03}{%
            \begin{ZX}[zx row sep=2.5ex, zx column sep=1em, ampersand replacement=\&]
    \zxN{} \rar \& \zxN{} \rar            \& \zxZ{}  \dar \rar \& \zxN{} \rar                 \& \zxN{} \\
                \&                        \& \zxZ{}            \&                             \&        \\
                \& \zxX{} \ar[ru] \ar[rd] \& \zxZ{\gamma} \dar \& \zxX{} \ar[lu] \ar[ld]      \&        \\
    \zxN{} \rar \& \zxX{} \uar            \& \zxZ{}            \& \zxX{} \uar \rar            \& \zxN{}
\end{ZX}
        }
    }}
    \\[2ex]
    =%
    \vcenter{\hbox{%
        \externalizezx{appendixphasegadgetdiagram04}{%
            \begin{ZX}[zx row sep=2.5ex, zx column sep=1em, ampersand replacement=\&]
    \zxN{} \rar \& \zxN{} \rar       \& \zxN{} \rar      \& \zxZ{} \rar         \& \zxN{} \rar    \& \zxN{}      \\
                \& \zxZ{\gamma} \rar \& \zxZ{}{} \dar    \& \zxZ{} \uar         \&                \&             \\
                \&                   \& \zxX{}{} \ar[ur] \& \zxX{} \ar[ul] \uar \&                \&             \\
    \zxN{} \rar \& \zxX{} \ar[ur]    \&                  \&                     \& \zxX{} \ar[ul] \& \zxN{} \lar \\
\end{ZX}
        }
    }}
    & \stackrel{\bialgrule}{=} %
    \vcenter{\hbox{%
        \externalizezx{appendixphasegadgetdiagram05}{%
            \begin{ZX}[zx row sep=2.5ex, zx column sep=1em, ampersand replacement=\&]
    \zxN{} \rar \& \zxN{} \rar       \& \zxZ{} \rar         \& \zxN{} \rar    \& \zxN{}      \\
                \& \zxZ{\gamma} \rar \& \zxX{} \uar         \&                \&             \\
                \&                   \& \zxZ{} \uar         \&                \&             \\
    \zxN{} \rar \& \zxX{} \ar[ur]    \&                     \& \zxX{} \ar[ul] \& \zxN{} \lar \\
\end{ZX}

        }
    }}
    \\[2ex]
    \stackrel{\idrule}{=} %
    \vcenter{\hbox{%
        \externalizezx{appendixphasegadgetdiagram06}{%
            \begin{ZX}[zx row sep=2.5ex, zx column sep=1em, ampersand replacement=\&]
    \zxN{} \rar \& \zxN{} \rar       \& \zxZ{} \rar      \&[3em] \zxN{} \\
                \& \zxZ{\gamma} \rar \& \zxX{} \uar      \&[3em]         \\
    \zxN{} \rar \& \zxN{} \rar       \& \zxZ{} \uar \rar \&[3em] \zxN{} \\
\end{ZX}
        }
    }}
    &
\end{align*}
\subsubsection{Proof of \eqref{eqn:phase-gadget-to-mbqc}}
\label{sec:proof_phase_gadet_to_mbqc}
\begin{align*}
    \vcenter{\hbox{%
        \externalizezx{appendixphasegadgetdiagram07}{%
            \begin{ZX}[row sep={2.5ex}, column sep=1em, ampersand replacement=\&]
    \zxN{} \rar \& \zxZ{} \rar      \&[1em] \zxN{} \rar       \& \zxN{}\\
                \& \zxX{} \uar      \&[1em] \zxZ{\gamma} \lar \&       \\
    \zxN{} \rar \& \zxZ{} \uar \rar \&[1em] \zxN{} \rar       \& \zxN{}\\
\end{ZX}

        }
    }}
    & \stackrel{\hadamardrule}{=} %
    \vcenter{\hbox{%
        \externalizezx{appendixphasegadgetdiagram08}{%
            \begin{ZX}[zx column sep=1em, ampersand replacement=\&]
    \zxN{} \rar \&[2em] \zxN{} \rar \& \zxZ{} \rar      \& \zxN{} \rar   \& \zxN{} \rar   \& \zxN{} \\
                \&[2em]             \& \zxH{} \uar      \&               \&               \&        \\
                \&[2em]             \& \zxZ{} \uar \rar \& \zxH{} \rar   \& \zxZ{\gamma}  \&        \\
                \&[2em]             \& \zxH{} \uar      \&               \&               \&        \\
    \zxN{} \rar \&[2em] \zxN{} \rar \& \zxZ{} \uar \rar \& \zxN{} \rar   \& \zxN{} \rar   \& \zxN{} \\
\end{ZX}

        }
    }}
    \\[2ex]
    & \stackrel{\substack{\hadamardrule \\ \spiderrule}}{=} %
    \vcenter{\hbox{%
        \externalizezx{appendixphasegadgetdiagram09}{%
            \begin{ZX}[zx column sep=1em, ampersand replacement=\&]
    \zxN{} \rar \&[2em] \zxN{} \rar \& \zxZ{} \rar      \& \zxN{} \rar   \& \zxN{} \\
                \&[2em]             \& \zxH{} \uar      \&               \&        \\
                \&[2em] \zxZ{} \rar \& \zxZ{} \uar \rar \& \zxX{\gamma}  \&        \\
                \&[2em]             \& \zxH{} \uar      \&               \&        \\
    \zxN{} \rar \&[2em] \zxN{} \rar \& \zxZ{} \uar \rar \& \zxN{} \rar   \& \zxN{} \\
\end{ZX}

        }
    }}
    \\[2ex]
    & \stackrel{\spiderrule}{=} %
    \vcenter{\hbox{%
        \externalizezx{appendixphasegadgetdiagram10}{%
            \begin{ZX}[zx column sep=1em, ampersand replacement=\&]
    \zxN{} \rar \&[2em] \zxN{} \rar \& \zxZ{} \rar      \& \zxN{} \rar       \& \zxN{} \\
                \&[2em]             \& \zxH{} \uar      \&                   \&        \\
                \&[2em] \zxZ{} \rar \& \zxZ{} \uar \rar \& \zxX{\gamma} \rar \& \zxX{} \\
                \&[2em]             \& \zxH{} \uar      \&                   \&        \\
    \zxN{} \rar \&[2em] \zxN{} \rar \& \zxZ{} \uar \rar \& \zxN{} \rar       \& \zxN{} \\
\end{ZX}

        }
    }}
    \\[2ex]
    &=%
    \vcenter{\hbox{%
        \externalizezx{appendixphasegadgetdiagram11}{%
            \begin{ZX}[zx column sep=1em, ampersand replacement=\&]
    \zxN{} \rar \&[2em] \zxN{} \rar \& \zxZ{} \rar      \& \zxN{} \rar       \& \zxN{} \\
                \&[2em]             \& \zxH{} \uar      \&                   \&        \\
                \&[2em] \zxZ{} \rar \& \zxZ{} \uar \rar \& \zxX{\gamma} \rar \& \zxX{2m\pi} \\
                \&[2em]             \& \zxH{} \uar      \&                   \&        \\
    \zxN{} \rar \&[2em] \zxN{} \rar \& \zxZ{} \uar \rar \& \zxN{} \rar       \& \zxN{} \\
\end{ZX}

        }
    }}
    \\[2ex]
    & \stackrel{\spiderrule}{=} %
    \vcenter{\hbox{%
        \externalizezx{appendixphasegadgetdiagram12}{%
            \begin{ZX}[zx column sep=1em, ampersand replacement=\&]
    \zxN{} \rar \&[2em] \zxN{} \rar \& \zxZ{} \rar      \& \zxN{} \rar       \& \zxN{}          \& \zxN{}      \\
                \&[2em]             \& \zxH{} \uar      \&                   \&                 \&             \\
                \&[2em] \zxZ{} \rar \& \zxZ{} \uar \rar \& \zxX{\gamma} \rar \& \zxX{m\pi} \rar \& \zxX{m\pi} \\
                \&[2em]             \& \zxH{} \uar      \&                   \&                 \&             \\
    \zxN{} \rar \&[2em] \zxN{} \rar \& \zxZ{} \uar \rar \& \zxN{} \rar       \& \zxN{}          \& \zxN{}      \\
\end{ZX}

        }
    }}
    \\[2ex]
    & \stackrel{\spiderrule}{=} %
    \vcenter{\hbox{%
        \externalizezx{appendixphasegadgetdiagram13}{%
            \begin{ZX}[zx column sep=1em, ampersand replacement=\&]
    \zxN{} \rar \&[2em] \zxN{} \rar \& \zxZ{} \rar      \& \zxN{} \rar       \& \zxN{}          \& \zxN{}      \\
                \&[2em]             \& \zxH{} \uar      \&                   \&                 \&             \\
                \&[2em] \zxZ{} \rar \& \zxZ{} \uar \rar \& \zxX{m\pi} \rar   \& \zxX{\gamma} \rar \& \zxX{m\pi} \\
                \&[2em]             \& \zxH{} \uar      \&                   \&                 \&             \\
    \zxN{} \rar \&[2em] \zxN{} \rar \& \zxZ{} \uar \rar \& \zxN{} \rar       \& \zxN{}          \& \zxN{}      \\
\end{ZX}

        }
    }}
    \\[2ex]
    & \stackrel{\picopyrule}{=} %
    \vcenter{\hbox{%
        \externalizezx{appendixphasegadgetdiagram14}{%
            \begin{ZX}[zx column sep=1em, ampersand replacement=\&]
    \zxN{} \rar \&[2em] \zxN{} \rar \& \zxZ{}       \rar      \& \zxN{} \rar       \& \zxN{}            \& \zxN{}      \\
                \&[2em]             \& \zxH{}       \uar      \&                   \&                   \&             \\
                \&[2em]             \& \zxX{m\pi}   \uar      \&                   \&                   \&             \\
                \&[2em] \zxZ{} \rar \& \zxZ{}       \uar \rar \& \zxN{} \rar       \& \zxX{\gamma} \rar \& \zxX{m\pi} \\
                \&[2em]             \& \zxX{m\pi}   \uar      \&                   \&                   \&             \\
                \&[2em]             \& \zxH{}       \uar      \&                   \&                   \&             \\
    \zxN{} \rar \&[2em] \zxN{} \rar \& \zxZ{}       \uar \rar \& \zxN{} \rar       \& \zxN{}            \& \zxN{}      \\
\end{ZX}
        }
    }}
    \\[2ex]
    & \stackrel{\hadamardrule}{=} %
    \vcenter{\hbox{%
        \externalizezx{appendixphasegadgetdiagram15}{%
            \begin{ZX}[zx column sep=1em, ampersand replacement=\&]
    \zxN{} \rar \&[2em] \zxN{} \rar \& \zxZ{}       \rar      \& \zxN{} \rar       \& \zxN{}            \& \zxN{}      \\
                \&[2em]             \& \zxZ{m\pi}   \uar      \&                   \&                   \&             \\
                \&[2em]             \& \zxH{}       \uar      \&                   \&                   \&             \\
                \&[2em] \zxZ{} \rar \& \zxZ{}       \uar \rar \& \zxN{} \rar       \& \zxX{\gamma} \rar \& \zxX{m\pi} \\
                \&[2em]             \& \zxH{}       \uar      \&                   \&                   \&             \\
                \&[2em]             \& \zxZ{m\pi}   \uar      \&                   \&                   \&             \\
    \zxN{} \rar \&[2em] \zxN{} \rar \& \zxZ{}       \uar \rar \& \zxN{} \rar       \& \zxN{}            \& \zxN{}      \\
\end{ZX}
        }
    }}
    \\[2ex]
    & \stackrel{\spiderrule}{=} %
    \vcenter{\hbox{%
        \externalizezx{appendixphasegadgetdiagram16}{%
            \begin{ZX}[zx column sep=1em, ampersand replacement=\&]
    \zxN{} \rar \&[2em] \zxN{} \rar \& \zxZ{}       \rar      \& \zxN{} \rar       \& \zxZ{m\pi} \rar       \& \zxN{}      \\
                \&[2em]             \& \zxH{}       \uar      \&                   \&                       \&             \\
                \&[2em] \zxZ{} \rar \& \zxZ{}       \uar \rar \& \zxN{} \rar       \& \zxX{\gamma} \rar     \& \zxX{m\pi} \\
                \&[2em]             \& \zxH{}       \uar      \&                   \&                       \&             \\
    \zxN{} \rar \&[2em] \zxN{} \rar \& \zxZ{}       \uar \rar \& \zxN{} \rar       \& \zxZ{m\pi} \rar       \& \zxN{}      \\
\end{ZX}
        }
    }}
\end{align*}

\subsection{Single-Qubit $Z$ Rotations in MBQC}
\label{sec:proof_zrotation_in_mbqc}
We show the conversion of single qubit Z-rotation into the MBQC paradigm by proofing \eqref{eqn:zrotation_in_mbqc}
\begin{align*}
    \vcenter{\hbox{%
        \externalizezx{appendixzrotationdiagram00}{%
            \begin{tikzcd}[thin lines, column sep=1em, row sep={4.0ex,between origins}, ampersand replacement=\&]
    \qw \&  \gate{R_Z(\gamma)} \& \qw 
\end{tikzcd}
        }
    }} \notag
    &=%
    \vcenter{\hbox{%
        \externalizezx{appendixzrotationdiagram01}{%
            \begin{ZX}[ampersand replacement=\&]
    \zxN{} \rar \&[1em] \zxZ{\gamma} \rar \&[1em] \zxN{}
\end{ZX}
        }
    }}\notag
    \stackrel{\spiderrule}{=} %
    \vcenter{\hbox{%
        \externalizezx{appendixzrotationdiagram02}{%
            \begin{ZX}[ampersand replacement=\&]
    \zxN{}      \&[1em] \zxZ{\gamma}     \&[1em] \zxN{} \\
    \zxN{} \rar \&[1em] \zxZ{} \rar \uar \&[1em] \zxN{}
\end{ZX}

        }
    }} %
    \\[2ex]
    & \stackrel{\hhrule}{=} %
    \vcenter{\hbox{%
        \externalizezx{appendixzrotationdiagram03}{%
            \begin{ZX}[ampersand replacement=\&]
    \zxN{}      \& \zxZ{\gamma} \dar     \&[1em] \zxN{}            \& \zxN{}      \&[0.5em] \zxN{}  \\
    \zxN{} \rar \& \zxZ{} \rar \ar[r, H] \&[1em] \zxN{} \ar[r, H]  \& \zxN{} \rar \&[0.5em] \zxN{} 
\end{ZX}

        }
    }} %
    \stackrel{\substack{\idrule \\ \spiderrule}}{=} %
    \vcenter{\hbox{%
        \externalizezx{appendixzrotationdiagram04}{%
            \begin{ZX}[ampersand replacement=\&]
    \zxN{}      \& \zxZ{\gamma} \dar     \&[1.0em] \zxN{}          \& \zxN{}               \&[1.0em] \zxN{}  \\
    \zxN{} \rar \& \zxZ{} \rar \ar[r, H] \&[1.0em] \zxX{m\pi} \rar \& \zxX{m\pi} \ar[r, H] \&[1.0em] \zxN{} 
\end{ZX}
        }
    }} %
    \\[2ex]
    &\stackrel{\picopyrule}{=} %
    \vcenter{\hbox{%
        \externalizezx{appendixzrotationdiagram05}{%
            \begin{ZX}[ampersand replacement=\&]
    \zxN{}      \& \zxZ{\gamma} \dar     \& \zxN{}               \&[1em] \zxN{}               \&[1.0em] \zxN{}  \\
    \zxN{} \rar \& \zxZ{} \rar           \& \zxZ{m\pi} \ar[r, H] \&[1em] \zxX{m\pi} \ar[r, H] \&[1.0em] \zxN{} 
\end{ZX}
        }
    }} %
    \\[2ex]
    & \stackrel{\fusionrule}{=} %
    \vcenter{\hbox{%
        \externalizezx{appendixzrotationdiagram06}{%
            \begin{ZX}[ampersand replacement=\&]
    \zxN{}      \& \zxZ{\gamma} \dar \rar \&[1em] \zxZ{m\pi}  \&                          \&[1.0em] \zxN{}  \\
    \zxN{} \rar \& \zxZ{} \rar  \ar[r, H] \&[1em] \zxZ{} \rar \& \zxX{m\pi} \ar[r, H]     \&[1.0em] \zxN{}  
\end{ZX}
        }
    }} %
    \\[2ex]
    & = %
    \vcenter{\hbox{%
        \externalizezx{appendixzrotationdiagram07}{%
            \begin{ZX}[ampersand replacement=\&]
    \zxN{} \rar \& \zxZ{\gamma} \ar[d, H] \rar \& \zxZ{m\pi}              \&[1.0em] \zxN{}  \\[3ex]
    \zxN{}      \& \zxZ{} \rar                 \& \zxX{m\pi} \ar[r, H]    \&[1.0em] \zxN{}  
\end{ZX}

        }
    }} %
    \stackrel{\fusionrule}{=} %
    \vcenter{\hbox{%
        \externalizezx{appendixzrotationdiagram08}{%
            \begin{ZX}[ampersand replacement=\&]
    \zxN{} \rar \& \zxN{} \rar \& \zxZ{} \ar[d, H] \rar \& \zxZ{\gamma}    \rar \& \zxZ{m\pi}      \& \zxN{} \\[3ex]
    \zxN{}      \& \zxZ{} \rar \& \zxZ{} \rar           \& \zxX{m\pi} \ar[r, H] \& \zxN{}     \rar \& \zxN{}  
\end{ZX}

        }
    }} %
    \\[2ex]
    & \stackrel{\hadamardrule}{=} %
    \vcenter{\hbox{%
        \externalizezx{appendixzrotationdiagram09}{%
            \begin{ZX}[ampersand replacement=\&]
    \zxN{} \rar \& \zxN{} \rar \& \zxZ{} \ar[d, H] \rar \& \zxZ{\gamma}    \rar \& \zxZ{m\pi}      \& \zxN{} \\[3ex]
    \zxN{}      \& \zxZ{} \rar \& \zxZ{} \rar           \& \zxZ{m\pi} \ar[r]    \& \zxN{}     \rar \& \zxN{}  
\end{ZX}
        }
    }} %
\end{align*}
\subsection{Single-Qubit $X$ Rotations in MBQC}
\label{sec:proof_mixer_in_mbqc}
We show the conversion of single qubit X-rotation into the MBQC paradigm by proofing \eqref{eqn:mixer-to-mbqc}
\begin{align*}
    &
    \vcenter{\hbox{%
        \externalizezx{appendixmixerdiagram00}{%
            \tikzstyle{basicshadow}=[blur shadow={shadow blur steps=8, shadow xshift=0.7pt, shadow yshift=-0.7pt, shadow scale=1.02}]
\tikzstyle{basic}=[draw,fill=white,basicshadow]
\tikzstyle{operator}=[basic,minimum size=1.5em]

\begin{tikzcd}[thin lines, column sep=1em, row sep={4.0ex,between origins}, ampersand replacement=\&]
    \qw \&  \gate{R_X(\beta)} \& \qw 
\end{tikzcd}
        }
    }} \notag
    \\[2ex]
    &=%
    \vcenter{\hbox{%
        \externalizezx{appendixmixerdiagram01}{%
            \begin{ZX}[ampersand replacement=\&]
    \zxN{} \rar \&[1em] \zxX{\beta} \rar \&[1em] \zxN{}
\end{ZX}
        }
    }}\notag
    \\[2ex]
    & \stackrel{\substack{\idrule \\ \spiderrule}}{=} %
    \vcenter{\hbox{%
        \externalizezx{appendixmixerdiagram02}{%
            \begin{ZX}[ampersand replacement=\&]
    \zxN{} \rar \&[1em] \zxZ{m_v \pi} \rar \& \zxZ{m_v \pi} \rar \& \zxX{\beta}\rar \&[1em] \zxN{}
\end{ZX}

        }
    }} %
    \\[2ex]
    & \stackrel{\picopyrule}{=} %
    \vcenter{\hbox{%
        \externalizezx{appendixmixerdiagram03}{%
            \begin{ZX}[ampersand replacement=\&]
    \zxN{} \rar \&[1em] \zxZ{m_v \pi} \rar \& \zxX{(-1)^{m_v}\beta}\rar \& \zxZ{m_v \pi} \rar \&[1em] \zxN{}
\end{ZX}
        }
    }} %
    \\[2ex]
    & \stackrel{\substack{\idrule \\ \spiderrule}}{=} %
    \vcenter{\hbox{%
        \externalizezx{appendixmixerdiagram04}{%
            \begin{ZX}[ampersand replacement=\&]
    \zxN{} \rar \&[1em] \zxZ{m_v \pi} \rar \& \zxX{(-1)^{m_v}\beta}\rar \& \zxX{m'_v \pi} \rar \& \zxX{m'_v \pi} \rar \& \zxZ{m_v \pi} \rar \&[1em] \zxN{}
\end{ZX}
        }
    }} %
    \\[2ex]
    & \stackrel{\spiderrule}{=} %
    \vcenter{\hbox{%
        \externalizezx{appendixmixerdiagram05}{%
            \begin{ZX}[ampersand replacement=\&]
                \&[1em]                    \& \zxX{(-1)^{m_v}\beta + m'_v \pi} \&                     \&                    \&[1em] \zxN{} \\[2ex] 
    \zxN{} \rar \&[1em] \zxZ{m_v \pi} \rar \& \zxX{} \rar \uar                 \& \zxX{m'_v \pi} \rar \& \zxZ{m_v \pi} \rar \&[1em] \zxN{}
\end{ZX}
        }
    }} %
    \\[2ex]
    & \stackrel{\hadamardrule}{=} %
    \vcenter{\hbox{%
        \externalizezx{appendixmixerdiagram06}{%
            \begin{ZX}[ampersand replacement=\&]
                \&[1em]                         \& \zxX{(-1)^{m_v}\beta + m'_v \pi} \&                     \&                    \&[1em] \zxN{} \\[2ex] 
    \zxN{} \rar \&[1em] \zxZ{m_v \pi} \ar[r, H] \& \zxZ{} \ar[r, H] \ar[u, H]       \& \zxX{m'_v \pi} \rar \& \zxZ{m_v \pi} \rar \&[1em] \zxN{}
\end{ZX}
        }
    }} %
    \\[2ex]
    & \stackrel{\spiderrule}{=} %
    \vcenter{\hbox{%
        \externalizezx{appendixmixerdiagram07}{%
            \begin{ZX}[ampersand replacement=\&]
                \&[0.5em] \zxZ{m_v \pi}         \& \zxX{(-1)^{m_v}\beta + m'_v \pi} \&             \&                    \&                    \&[1em] \zxN{} \\[2ex] 
    \zxN{} \rar \&[0.5em] \zxZ{} \uar \ar[r, H] \& \zxZ{} \ar[r, H] \ar[u, H]       \& \zxZ{} \rar \&\zxX{m'_v \pi} \rar \& \zxZ{m_v \pi} \rar \&[1em] \zxN{}
\end{ZX}
        }
    }} %
    \\[2ex]
    & \stackrel{\hadamardrule}{=} %
    \vcenter{\hbox{%
        \externalizezx{appendixmixerdiagram08}{%
            \begin{ZX}[ampersand replacement=\&]
                \&[0.5em] \zxZ{m_v \pi}         \& \zxZ{(-1)^{m_v}\beta + m'_v \pi} \&             \&                    \&                    \&[1em] \zxN{} \\[2ex] 
    \zxN{} \rar \&[0.5em] \zxZ{} \uar \ar[r, H] \& \zxZ{} \ar[r, H] \ar[u]          \& \zxZ{} \rar \&\zxX{m'_v \pi} \rar \& \zxZ{m_v \pi} \rar \&[1em] \zxN{}
\end{ZX}
        }
    }} %
    \\[2ex]
    & \stackrel{\spiderrule}{=} %
    \vcenter{\hbox{%
        \externalizezx{appendixmixerdiagram09}{%
            \begin{ZX}[zx row sep=4ex, zx column sep=1em, ampersand replacement=\&]
    \zxN{} \rar \&[2em] \zxN{} \rar \& \zxZ{}  \rar           \& \zxZ{m_v \pi}                \&                      \&        \\
                \&[2em] \zxZ{} \rar \& \zxZ{}  \ar[u, H] \rar \& \zxZ{(-1)^{m_v}\beta} \rar   \& \zxZ{m'_v \pi}       \&        \\
                \&[2em] \zxZ{} \rar \& \zxZ{}  \ar[u, H] \rar \& \zxX{m'_v \pi} \rar          \&  \zxZ{m_v \pi} \rar \& \zxN{} \\
\end{ZX}
        }
    }} %
\end{align*}

\subsection{QAOA for QUBO Problems in MBQC}
\label{sec:proof_qaoa_mbqc_qubo}
We derive the MBQC formulation of QAOA for the general QUBO case \eqref{eqn:qaoa_mbqc_qubo}.
As in Section~\ref{sec:qaoa_mbqc}, we begin be considering the contribution of a single edge $(u, v)$ and a single QAOA layer. 
Again, for the benefit of the reader we visually mark the relevant spiders with blue and red dots.
\begin{align*}
    &
    \vcenter{\hbox{%
        \externalizezx{appendixqaoambqcqubo00}{%

        }
    }} \notag
\end{align*}

\label{sec:appendix}

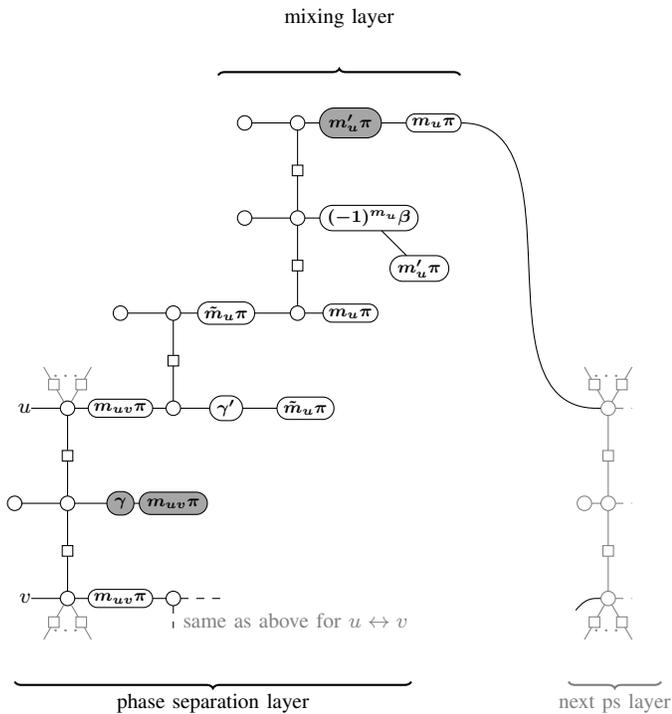
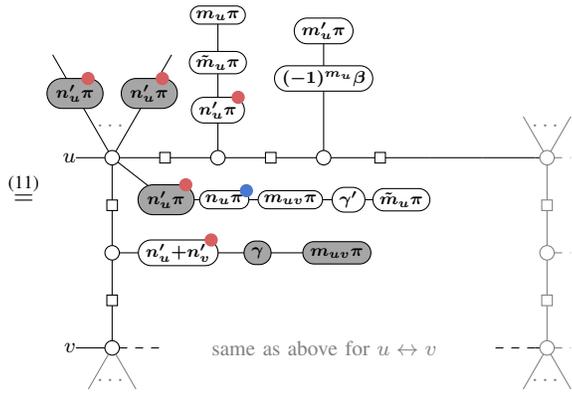
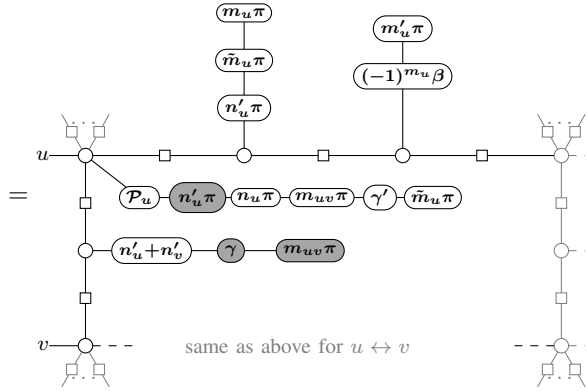
\end{document}